 \documentclass[12pt,preprint]{aastex}

\shorttitle{Stellar Parameters and Metallicities}
\shortauthors{Ghezzi et al.}

\begin{document}


\title{Stellar Parameters and Metallicities of Stars Hosting Jovian and Neptunian mass
Planets: A Possible Dependance of Planetary Mass on Metallicity\footnote{Based on observations 
made with the 2.2 m telescope at the European Southern Observatory (La Silla, Chile), 
under the agreement ESO-Observat\'orio Nacional/MCT.}}


\author{L. Ghezzi\altaffilmark{1}, K. Cunha\altaffilmark{1,2,3}, V. V. Smith\altaffilmark{2}, F. X. de Ara\'ujo\altaffilmark{1,4},  S. C. Schuler\altaffilmark{2} \& R. de la Reza\altaffilmark{1}}


\altaffiltext{1}{Observat\'orio Nacional, Rua General Jos\'e Cristino, 77, 20921-400, 
                 S\~ao Crist\'ov\~ao, Rio de Janeiro, RJ, Brazil; luan@on.br}
\altaffiltext{2}{National Optical Astronomy Observatory, 950 North Cherry Avenue, Tucson, AZ 85719, USA}
\altaffiltext{3}{Steward Observatory, University of Arizona, Tucson, AZ 85121, USA}
\altaffiltext{4}{Deceased in July 2009}


\begin{abstract}
The metal content of planet hosting stars is an important ingredient which
may affect the formation and evolution of planetary systems.
Accurate stellar abundances require the determinations of reliable physical
parameters, namely the effective temperature, surface gravity,
microturbulent velocity, and metallicity. This work presents the
homogeneous derivation of such parameters for a large sample of
stars hosting
planets (N=117), as well as a control sample of
disk stars not known to harbor giant, closely orbiting planets (N=145).
Stellar parameters and iron abundances are derived from an automated analysis technique
developed for this work.
As previously found in the literature, the results in this study indicate that
the metallicity distribution of planet hosting stars is more metal-rich
by $\sim$0.15 dex when compared to the control sample stars. A segregation
of the sample according to planet mass indicates that the metallicity
distribution of stars hosting only Neptunian-mass planets (with no Jovian-mass planets)
tends to be more metal-poor in comparison with that obtained for stars hosting a closely
orbiting Jovian planet. The significance of this difference in metallicity
arises from a homogeneous analysis of samples of FGK dwarfs which do not include
the cooler and more problematic M dwarfs.
This result would indicate that there is a possible link between planet mass
and metallicity such that metallicity plays a role in setting the mass of the
most massive planet. Further confirmation, however, must await larger samples.
\end{abstract}





\section{Introduction}

\label{int}

More than 380 stars with planets have been discovered to date, half of which were
detected in
the past three years. Most of the extrasolar planets have been discovered via
radial-velocity
measurements of the reflex motions of the planet-hosting star and such surveys
are biased to detect preferentially the largest and most closely orbiting planets.
Within an ever increasing sample size,  one statistically significant
property has been confirmed for these objects: the average metallicity of the
solar-like stars known to have giant planets (i.e. those planets close to the mass of
Jupiter
or larger) is higher when compared to field F, G and K dwarfs not known to host giant
planets (see, e.g., \citealt{g97}; \citealt{santos01}; \citealt{l03}; \citealt{s05}; \citealt{fv05};
\citealt{b06}; \citealt{s08}). This difference is attributed to two possible
scenarios: primordial enrichment and pollution. At present, the former seems to best
account for the metal-rich
nature of the planet hosting stars, since the probability of finding a planet is a
steeply rising function
of the stellar metallicity (e.g., \citealt{s04}; \citealt{fv05}). However, the pollution
hypothesis cannot
be discarded, as contradictory conclusions have been found by several studies which
attempted to unveil other
chemical peculiarities in planet hosting stars (for a comprehensive review, see
\citealt{g06}; \citealt{us07}).

In addition to the population of rather metal rich stars hosting Jovian-mass planets,
there is a growing number of known systems with considerably
lower-mass planetary companions. The range of planetary masses now
includes objects with minimum masses of only about
$M_{p}\sin i \sim 4M_{\earth}$, with many systems containing
\textquotedblleft Neptunian-mass\textquotedblright\ planets, with $M_{p}\sin i< 25M_{\earth}$.
It is of interest to investigate whether the trend for Jovian-mass
planets to have a metal-rich stellar parent continues towards systems
with lower-mass planets that do not contain the large Jovian-mass
planets. \citet{u06}, \citet{s08}, and \citet{mayor09} 
suggest that stars which have as their most massive planets
Neptunian-mass objects may not be metal rich; however, the number of
such systems which have been studied is just a few. The list of
stars with Neptunian-mass planets continues to grow and these objects
will help to probe the possibility of a stellar-metallicity planet-mass
connection.

The observed variety of exo-planetary masses and orbital separations, along
with evidence of planetary migration and its possible influence on proto-planetary disk
-- stellar
interactions suggests that it is of importance to determine chemical abundance
distributions in different populations of exo-planet host stars.
The search for subtle patterns in the abundances of stars with and without planets
that may reveal details of planetary formation or planetary system architecture
is based ideally on a homogeneous and self-consistent analysis.
If all samples are observed with the same instrumental setup and analyzed with
a consistent methodology, systematic effects are more likely to be avoided.
This study sets forth a homogeneous determination of stellar parameters and
metallicities for a large sample of stars with planets, including a few stars hosting
only Neptunian mass planets, as well as a control sample comprised of field stars
not known to host giant planets. \S \ref{data}
describes the observational data, sample selection criteria, and data reduction.
The determination of stellar parameters, effective temperatures, surface gravities and
metallicities,
including the adopted iron line list, are presented in \S \ref{analysis}.
In \S \ref{disc}, results from this work are compared with those from the literature and
discussed in light of various planet-metallicity correlations.
Included in this discussion is an investigation into whether metallicity plays a role in
determining the
mass of the most massive planet in an exo-planetary system.
Finally, concluding remarks are presented in \S \ref{conc}.
The derivation of the elemental abundances other than Fe will be treated in subsequent
papers.


\section{Observations and Data Reduction}

\label{data}

\subsection{Sample Selection}

\label{sample}

{\bf Planet Hosting Stars}:
The sample of main-sequence stars with planets analyzed in this study contains 117 targets. The target list was compiled 
using the Extrasolar Planet Encyclopaedia\footnote{Available at http://exoplanet.eu} and 
updated with newly discovered systems until August 2008. We selected all planet
hosting stars having $\delta < +26\degr$  and $V < 12$.
The declination limit was imposed by object observability at La Silla Observatory in Chile,
and the limiting magnitude was set in order to keep exposure times needed to achieve the desired S/N
relatively short. Several stars in this sample have been previously analyzed in recent studies
of planet hosting stars (\citealt{l03}; \citealt{s04,s05}; \citealt{t05}; \citealt{vf05};
\citealt{lh06}; \citealt{b06}; \citealt{s08}). We note that 16 planet hosting stars in this sample
are not included in these previous abundance studies.

{\bf Control Sample}:
A control sample of main-sequence disk stars which are not known to host giant planets was selected from 
the list of nearby F, G and K stars in \citet{fv05} which has been targeted in the planet 
search programs conducted at the Keck Observatory, Lick Observatory and the Anglo-Australian Telescope.
That study identified 850 stars for which there are enough observations 
to securely detect the presence of companions with velocity amplitudes $K >$ 30 m $\rm s^{-1}$ and orbital periods shorter 
than 4 yr. From the subsample of stars with non-detections of giant planets, 
we  eliminated stars with [M/H] $<$ -1.0; $v$ sin $i >$ 10 km $\rm s^{-1}$ 
(typical rotational velocities are much lower for solar-type stars) and $\delta > +26\degr$. 
In addition, any stars which were found subsequently to host giant planets (the only case being HD 16417) 
were obviously removed from the list. 
Binaries, as well as targets having one single spectrum analyzed were
also excluded (according to Table 8 in \citealt{vf05}). HD 36435 was 
added to the list as it was previously analyzed for $^{6}$Li (\citealt{ghezzi09}).
The final sample of comparison stars in this study has 145 targets. 
A list with all targets analyzed, planet hosting stars as well as control sample stars, 
is presented in Table \ref{obslog}.

\subsection{Observations and Data Reduction}

\label{obs}
High-resolution spectra were obtained with the Fiber-fed Extended Range Optical Spectrograph (FEROS; \citealt{kaufer99}) 
attached to the MPG/ESO-2.20m telescope (La Silla, Chile). The detector was a 2k X 4k EEV CCD with 
15 $\mu$m pixels. This instrumental setup produces spectra with almost complete spectral coverage 
from 3,560 to 9,200 \AA\ (over 39 \'echelle orders) and at a nominal resolution R $= \lambda/\Delta \lambda \sim$ 48,000.
The observations were conducted during 6 observing runs between April 2007 and August 
2008\footnote{Under the agreement ESO-Observat\'orio Nacional/MCT}.
A solar spectrum of the afternoon sky (T$_{exp}$= 2x120s) was taken before each observing night.
A detailed log of the observations, including $V$ magnitudes, observation dates, total integration times and the resulting S/N per 
resolution element, is found in Table \ref{obslog}.

The spectra were reduced with the FEROS Data Reduction System 
(DRS)\footnote{Available at http://www.eso.org/sci/facilities/lasilla/instruments/feros/tools/DRS/index.html}. 
The data reduction followed standard procedures. 
An average flat-field image was used in order to define the positions of the \'echelle orders. 
The background (bias level and scattered light) was subtracted from the images. 
The bias level was determined from the overscan region of the CCD and the scattered light was measured in 
the interorder space and in the region between the two fibers. 
The extraction of the \'echelle orders was done with a standard algorithm that also finds and removes cosmic rays. All extracted images were 
divided by the average flat-field in order to remove pixel-to-pixel variations and they were corrected for the blaze function. 
The flat-fielded spectra were wavelength calibrated using ThArNe and/or ThAr+Ne calibration frames. 
The calibrated spectra were rebinned in constant steps of wavelength and a barycentric correction was applied. 
Finally, the reduced spectra were corrected for radial velocity shifts by comparing the observed wavelengths 
of some isolated and moderately strong iron lines with their rest wavelenghts taken from Vienna Atomic Line 
Database\footnote{Available at http://ams.astro.univie.ac.at/$\sim$vald/} (VALD; \citealt{k99}).


\section{Analysis}

\label{analysis}

\subsection{The Fe Line List}

\label{list}

The line lists for \ion{Fe}{1} and \ion{Fe}{2} were compiled from the line sample in \citet{s08} and \citet{mb09}. 
The initial line list contained over 100 iron lines but 
using both the Solar Flux Atlas (\citealt{k84}) as well as the solar spectrum taken with FEROS on 20 August 2008, 
and from results of test calculations with a variety of \textit{gf}-values from the literature, we selected  
27 \ion{Fe}{1} and 12 \ion{Fe}{2} suitable lines which were unblended and of intermediate strength
(equivalent widths less than 90 m\AA, in order to limit the effects of damping on 
the abundance determination). 
The final line list adopted in this Fe abundance analysis is presented in Table \ref{linelist}. 
The wavelengths and lower excitation potentials (LEP) of the Fe transitions were taken from VALD. 
The \textit{gf}-values for \ion{Fe}{1} transitions were taken from: 
\citet{b82a,b82b,b84,b86,b95}, \citet{bkk91} and \citet {bk94}, and \citet{ob91}. The \textit{gf}-values 
from these studies were carefully compared in \citet{l96} and found to be in excellent agreement.
\citet{f06} also argue that differences between the 
\textit{gf}-values of these three groups are comparable to random uncertainties in the measurements.
Corrections to the log \textit{gf} scales from these different sources were therefore deemed not necessary in this study;
whenever a transition had more than one \textit{gf}-value available, an average value was adopted.
The \textit{gf}-values for the \ion{Fe}{2} lines in this study were taken from the critical analysis 
of \citet{l96}.


\subsection{Equivalent Width Measurements}

\label{ew}

The code ARES\footnote{http://www.astro.up.pt/~sousasag/ares/} 
(\citealt{s07}) was used in order to measure equivalent widths (EWs) of sample Fe lines automatically. 
Briefly, this program first fits a polynomial to the local continuum in a spectral region defined 
by the user. It then determines which lines inside the given interval can be fit by a Gaussian profile. 
Finally, it computes the equivalent width(s) for the line(s) of interest assuming a Gaussian profile. 
More details about the ARES code can be found in \citet{s07} and \citet{s08}. The equivalent widths measured 
for all program stars can be found in \citet{ghezzi10}.

\begin{figure}
\epsscale{1.00}
\plotone{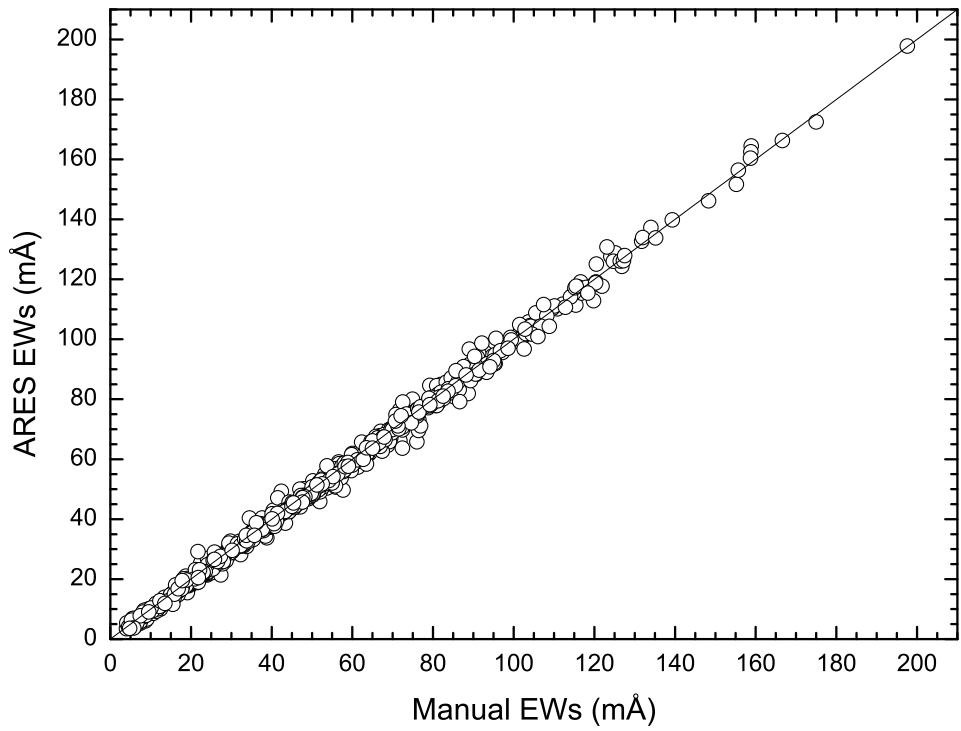}
\caption{A comparison between manual (using IRAF \texttt{splot}) and automatic measurements of equivalent widths (using ARES code) for a sample of \ion{Fe}{1} and \ion{Fe}{2} lines in six target stars and the Sun. The solid line represents perfect agreement.}
\label{ews}
\end{figure}

Possible systematic effects in the automatic ARES equivalent width measurements were investigated here. We 
measured (using the \texttt{splot} task of IRAF) equivalent widths of 75 \ion{Fe}{1} lines and 22 \ion{Fe}{2} lines in 6 stars which were selected to bracket the range in effective temperature, metallicity and spectrum quality (signal-to-noise ratio) of our sample as well as the Sun.
A comparison between manual
and automatic equivalent width measurements for 638 lines is shown in Figure \ref{ews}.
The mean difference between the two sets of equivalent widths is 
$\langle$EW$_{ARES}-$EW$_{Manual}\rangle = -0.53 \pm 2.10$ m\AA. Also, the following linear fit is obtained: 
EW$_{ARES} = (1.005 \pm 0.002)$EW$_{Manual} + (-0.77 \pm 0.16)$. The correlation coefficient and the standard 
deviation are R $=$ 0.99627 and $\sigma = 2.09$ m\AA, respectively.  


The exercise above indicates that the equivalent widths which were measured automatically using the ARES code
are consistent with our measurements, although there is a slight trend of ARES equivalent widths being marginally smaller than the ones measured manually in this study. 
This result is in line with what was found in \citet{s07}. Although we find an overall 
good agreement between manual and automatic equivalent widths, it is important to carefully 
check the results because ARES does not make quality assessments
of the measurements it outputs. For instance, in the tests decribed above we note that
there were 10 lines with obviously erroneous equivalent width measurements, which were discarded. 
In this study we estimate an uncertainty of $\sim$2 m\AA\ as the typical uncertainty in the equivalent width measurements.
Differences in equivalent widths of $\pm$ 2 m\AA\ are about what is expected given the resolution, sampling,
and S/N of the spectra and no significant systematic effects are found between ARES and manual measurements. 

\subsection{Derivation of Stellar Parameters and Iron Abundances}

\label{paratm}

Stellar parameters (T$_{eff}$, $\log g$ and $\xi$) and metallicities ([Fe/H]) were derived 
homogeneously and following standard spectroscopic methods which are based on requirements
of excitation and ionization equilibria. This abundance analysis was done in Local Thermodynamic Equilibrium (LTE) using 
the 2002 version of MOOG\footnote{Available at http://verdi.as.utexas.edu/moog.html.} (\citealt{s73}). 
In all calculations van der Waals constants were multiplied by an enhancement factor of 2.0 (\citealt{h91}).
The model atmospheres in this study were interpolated from the ODFNEW grid of 
ATLAS9 models\footnote{Available at http://kurucz.harvard.edu/} (\citealt{ck04}). 

\begin{figure}
\epsscale{0.70}
\plotone{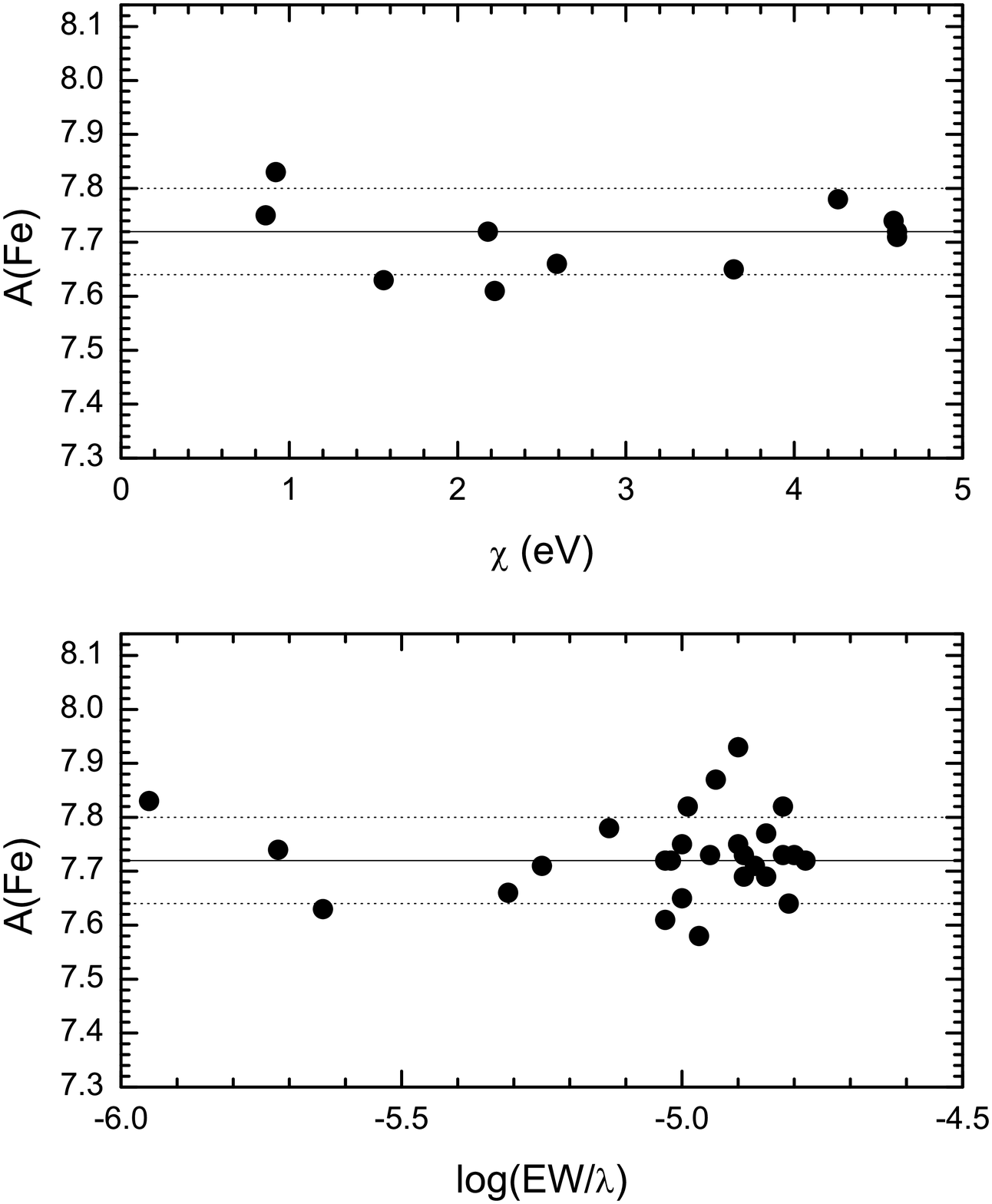}
\caption{The spectroscopic determination of effective temperature and microturbulent velocity for HD 2039 
obtained from zero slopes in the runs of \ion{Fe}{1} abundances with excitation potential of the transitions 
(top panel) and reduced equivalent widths (bottom panel). The solid line represents the mean iron abundance and
the dashed lines represent the 1-$\sigma$ of the distribution. The top panel shows only those lines with 
$\log (EW/\lambda) < -5.00$ which were used in the T$_{eff}$ iteration. 
\ion{Fe}{1} and \ion{Fe}{2} abundances are consistent  and the slopes are zero for A(Fe) = 7.73.}
\label{moog}
\end{figure}

Effective temperatures and microturbulent velocities were iterated until the slopes of 
A(\ion{Fe}{1})\footnote {A(\ion{Fe}{1}) = log [N(\ion{Fe}{1})/N(H)] + 12} \textit{versus} excitation potential, $\chi$, and A(\ion{Fe}{1}) \textit{versus} reduced equivalent
width,\linebreak $\log(EW/\lambda)$, were respectively zero (excitation equilibrium). 
Only lines with $\log (EW/\lambda) < -5.00$ (this limit was 
changed to larger values for cooler stars) were used in the first iteration, in order to decouple the 
T$_{eff}$ and $\xi$ determinations. Surface gravities were iterated until \ion{Fe}{1} 
and \ion{Fe}{2} returned the same mean abundances (ionization equilibrium). At the end of the iterative 
process, a consistent set of values of T$_{eff}$, log g and microturbulent velocity as well as
the mean \ion{Fe}{1} (= \ion{Fe}{2}) abundance is obtained for the star. This procedure was adopted
for all stars in our sample except for seven targets which had lower metallicities and solar temperatures 
(namely HD 6434, HD 51929, HD 80913, HD 114762, HD 153075, HD 155918 and HD 199288). In these cases,
the microturbulent velocities were kept at a fixed value because there were no lines strong enough in order to anchor 
the iteration of this parameter.
As an example, Figure \ref{moog} shows the final iterated plots of A(\ion{Fe}{1}) \textit{versus} $\chi$ 
(top panel) and A(\ion{Fe}{1}) \textit{versus} $\log (EW/\lambda)$ (bottom panel) for target HD 2039. 


{\bf An Automated Analysis}: Due to the large number of stars in our sample and Fe lines included
in the analysis,
BASH and FORTRAN codes were built in order to automate the whole iterative process described above. In summary,
the code starts with automatic equivalent width measurements using ARES (Section \ref{ew}) and iterates to a final 
set of consistent values of effective temperature, surface gravity, microturbulence and
Fe abundance (both from \ion{Fe}{1} and \ion{Fe}{2}). With the development of an automatic procedure 
it is now possible to analyze the entire sample of over 300 stars studied here in a few days without interventions.

In order to further test our line list (Table \ref{linelist}) and analysis method, the solar spectrum 
(observed with FEROS spectrograph
on August 20, 2008) was analyzed in a similar manner, with automatic measurements of equivalent widths 
(the measured solar equivalent widths are found in the last column of Table \ref{linelist}). 
Solar abundances A(\ion{Fe}{1}) = 7.43 $\pm$ 0.07 and A(\ion{Fe}{2}) = 7.44 $\pm$ 0.05 as well as a microturbulence
$\xi$ = 1.00 km $\rm s^{-1}$ were derived using a Kurucz ODFNEW model atmosphere with $T_{\rm eff}$ = 5777 K, 
$\log g$ = 4.44, a model
turbulence of $\xi$ = 2.0 km $\rm s^{-1}$, and l/H$_{p}$ = 1.25. This solar Fe abundance is in excellent agreement with 
the results of \citet{r03} and \citet{f06} (A(Fe)=7.45), which use \textit{gf}-values from the same sources as here. 
The derived solar iron abundance also compares well (within the uncertainties) with recent solar abundance determinations 
for 3D hydrodynamical models from \citet[A(Fe)= 7.50 $\pm$ 0.04]{a09} and \citet[A(Fe)=7.52 $\pm$ 0.06]{caf10}.

Final values of effective temperatures, surface gravities and microturbuleny velocities for all stars are presented in Table \ref{atm_par}. 
The metallicities [Fe/H] (listed in the last column in Table \ref{atm_par}) were calculated for a solar 
abundance A(Fe)$_{\sun}$ = 7.43 (as derived here). The $\sigma$-values listed in columns 6 and 8 correspond to 
the standard deviations of the final mean abundances of \ion{Fe}{1} and \ion{Fe}{2}. 
The number of \ion{Fe}{1} and \ion{Fe}{2} lines considered for each star are listed in columns 7 and 9, respectively.

\begin{figure}
\epsscale{1.00}
\plotone{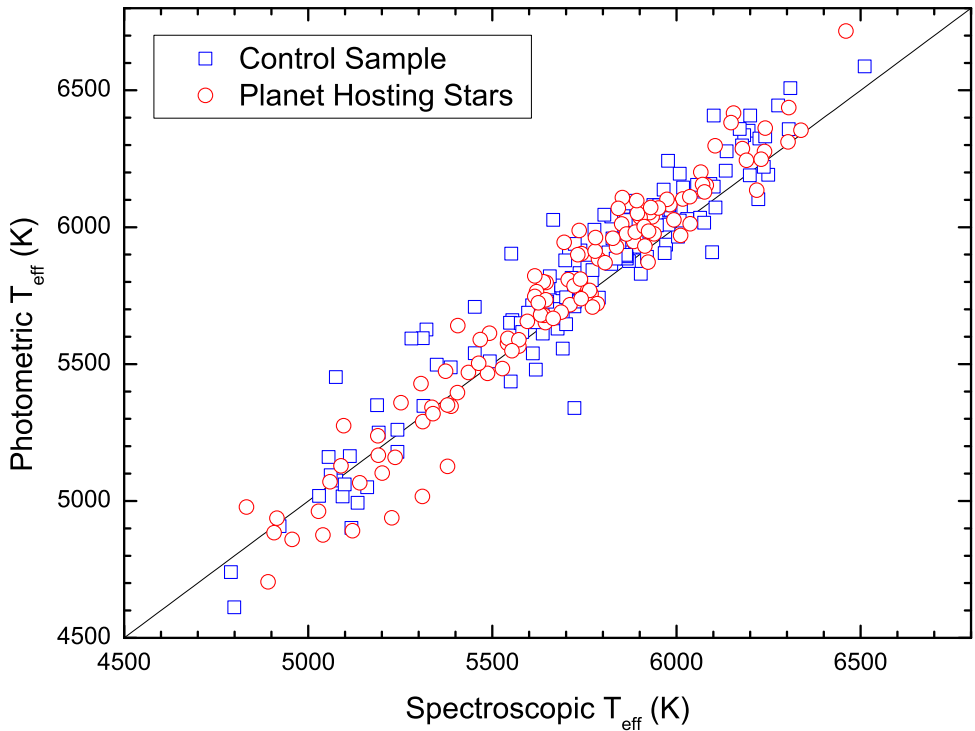}
\caption{A comparison between photometric temperatures derived with the $V-K$ calibration in \citet{c10}
and spectroscopic temperatures derived in this study for planet hosting stars (red open circles) and control 
sample (blue open squares). The solid line represents the bisector. The two effective temperature scales are in reasonable agreement, however, there is a tendency for the photometric $T_{eff}$s to be higher than the spectroscopic ones at the high $T_{eff}$ end and lower at the low $T_{eff}$ end.}
\label{tefs}
\end{figure}

As a comparison, photometric effective temperatures were also derived using the T$_{eff}$ \textit{versus} $V-K$ calibration recently
published by  \citet{c10}. The $V$ and $K_{s}$ magnitudes for the target stars were taken, respectively, from \textit{The Hipparcos and Tycho Catalogues} (\citealt{hip97}) and the \textit{2MASS} All-Sky Catalog of Point Sources 
(\citealt{c03}). A comparison between the spectroscopic and photometric temperatures shows that the latter 
are systematically higher (Figure \ref{tefs}).
The mean difference for all studied stars is $\Delta T_{eff}(spec-phot) = -63 \pm 113$ K, indicating reasonable
agreement.

Uncertainties in the parameters $T_{\rm eff}$, log g, $\xi$ and [Fe/H] were estimated as in \citet{gv98} and 
can be seen in Table \ref{atm_par}. We note that these are internal errors and that the real uncertainties 
might be somewhat larger.
Departures from LTE were not considered in this study and these can affect the derived LTE abundances.
Non-LTE effects are expected to be smaller for \ion{Fe}{2} lines as \ion{Fe}{2} (and not \ion{Fe}{1}) is
the dominant ionization stage in solar type stars. For \ion{Fe}{1}, departures from LTE are larger and may be
at the level of $\sim$0.1 dex (\citealt{g01a,g01b}).


\subsection{Evolutionary Parameters}

\label{evol}

Stellar luminosities, masses, radii and ages define the evolutionary stages of stars and for this
sample these were calculated in the following way. 
Absolute magnitudes $M_{V}$ were determined using the classical formula: 
\begin{equation}
 \label{Mv}
 M_{V}=V+5+5\log \pi-A_{V}.
\end{equation}
As already mentioned, the apparent $V$ magnitudes (Table \ref{obslog}) were taken from \textit{The Hipparcos and Tycho Catalogues} (\citealt{hip97}). As the 
uncertainties in this magnitude are not listed, we adopted the errors in $V_{T}$, given the similarities 
between the two passbands (\citealt{vL97}). The parallaxes $\pi$ and their uncertainties (Table \ref{evol_par}; columns 
2 and 3)  were taken from \citet{vL07}. Three stars 
(namely HD 70573, BD-10 3166 and 
WASP 2) were not present in the sources above, thus their $V$ magnitudes and parallaxes come 
from the references in SIMBAD\footnote{http://simbad.u-strasbg.fr/simbad/} and The Extrasolar Planet 
Encyclopaedia. The interstellar monochromatic extinction at $V$ magnitude, 
$A_{V}$ (Table \ref{evol_par}; column 4), as well as its error were calculated using the tables from \citet{a92} and the code EXTINCT.FOR from 
\citet{h97}. Note that the \citet{a92} model is accurate to distances within 1 kpc of the Sun. 

Absolute magnitudes were converted to bolometric magnitudes $M_{bol}$ by adding bolometric corrections 
in the $V$ band, $BC_{V}$ (Table \ref{evol_par}; column 5), linearly interpolated from the grids of \citet{g02} for the atmospheric parameters given 
in Table \ref{atm_par}. The luminosities were then calculated using the well known relation:
\begin{equation}
 \label{logL}
 \log\frac{L}{L_{\sun}}=-0.4(M_{bol}-M_{bol,\sun}),
\end{equation}
where $M_{bol,\sun}=4.77$ (\citealt{g02}). The uncertainties in $M_{V}$, $M_{bol}$ and $\log(L/L_{\sun})$ were 
derived considering that $\sigma(BC_{V})$ and $\sigma(M_{bol,\sun})$ 
are zero. The luminosities and uncertainties are listed in Table \ref{evol_par} (columns 6 and 7).

\begin{figure}
\epsscale{0.80}
\plotone{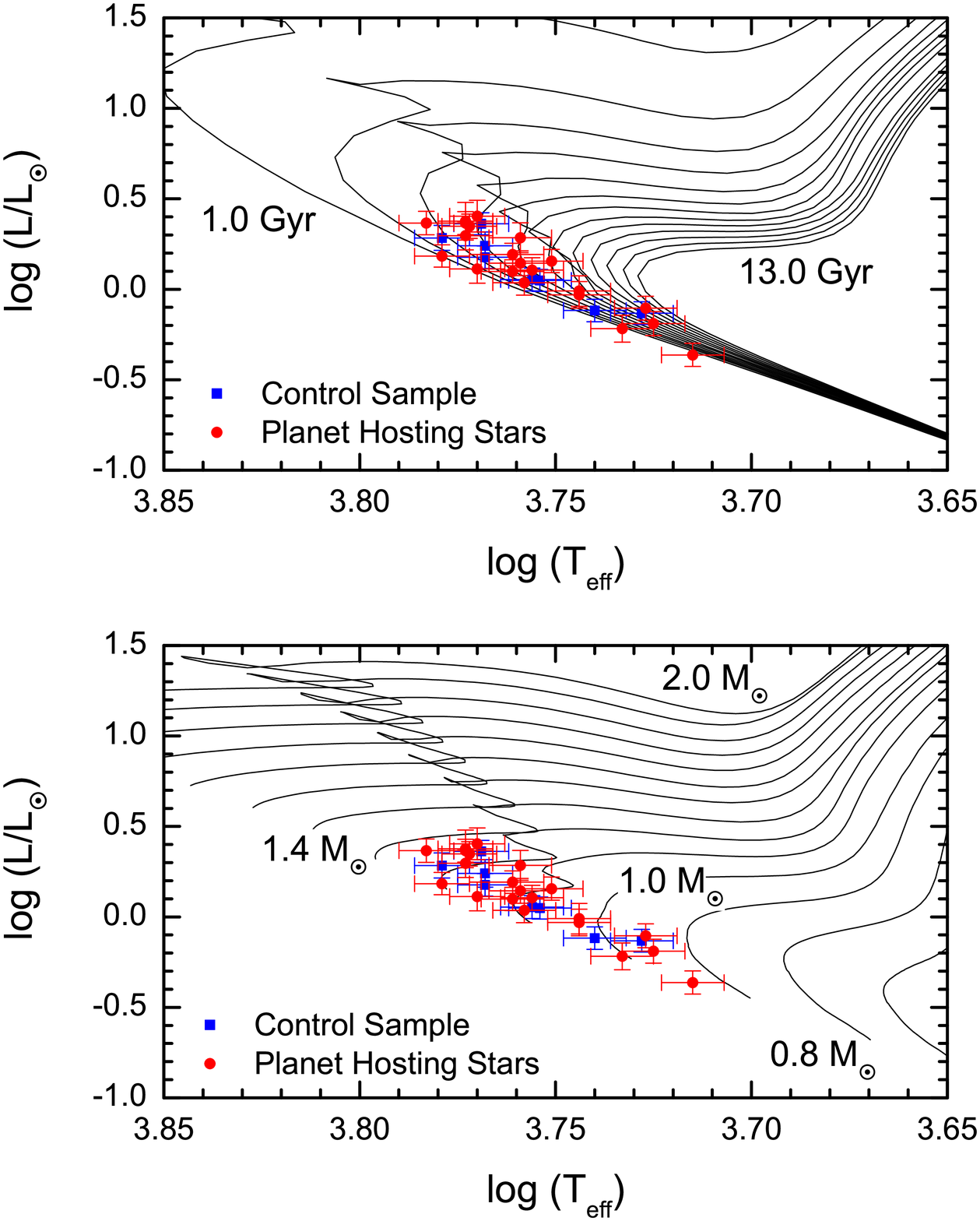}
\caption{The location of sample planet hosting stars (red circles) and
control disk sample stars (blue squares) in an HR diagram. The values of effective temperature and luminosity 
for the targets are from Tables \ref{atm_par} and \ref{evol_par}, respectively. 
The top panel shows isochrones for ages varying
between 1 and 13 Gyr and the bottom panel shows evolutionary tracks for mass tracks between 0.8 and 2 $M_{\sun}$.
The grids of isochrones and evolutionary tracks were calculated for [Fe/H] = +0.30 dex (\citealt{d04}; \citealt{y03})}
\label{age_mass}
\end{figure}

Effective temperatures and luminosities were used to place the stars on a grid of Y$^{2}$ isochrones from \citet{d04}, thus allowing for an age determination.
An interpolation code provided by the authors\footnote{Available at http://www.astro.yale.edu/demarque/yyiso.html} 
was used in order to obtain a set of isochrones ranging from 1.0 to 13.0 Gyr in age and from $-$0.70 dex to +0.40 dex in metallicity 
(with steps of 1.0 Gyr and 0.1 dex, respectively). 
As an example, we show the grid of isochrones for [Fe/H] = +0.30 dex in Figure \ref{age_mass} (top panel). 
The locations of all targets stars are indicated. The age of the closest isochrone was attributed for each star.
Given the uncertainties in T$_{eff}$, $\log(L/L_{\sun})$ and [Fe/H] and the proximity of the isochrones, 
an age interval was also estimated for each star together with a single age (see Table \ref{evol_par}). 
Also, some stars were located outside the grid and their ages were indicated as lower 
or upper limits. 

Stellar radii and spectroscopic masses M$_{spec}$ (as well as their uncertainties) were derived using 
standard relations (see e.g. eqs. 5-8 from \citealt{vf05}). 
The masses were also calculated by placing the stars on a grid of evolutionary tracks from \citet{y03}. 
The mass of the closest track was attributed for each star.
An interpolation code (provided by the authors\footnote{Available at http://www.astro.yale.edu/demarque/yystar.html}) 
was used in order to obtain a set of tracks ranging from 0.5 to 2.0 $M_{\sun}$ in mass and from $-$0.70 dex to +0.40 dex in metallicity (with steps of 0.1 $M_{\sun}$ and 0.1 dex, respectively). 
A typical uncertainty in mass of 0.1 
$M_{\sun}$ was estimated by considering the errors in T$_{eff}$, $\log(L/L_{\sun})$ and [Fe/H]. 
In some cases, however, this error was larger because of the location of the star on the grid 
or due to a larger uncertainty in the luminosity. Also, a few stars were located below the Zero Age Main Sequence (ZAMS) and their 
masses had to be estimated through an extrapolation. As an example, the grid of evolutionary tracks for 
[Fe/H] = +0.30 dex is shown in Figure \ref{age_mass} (bottom panel). The stellar radii and masses 
(spectroscopic, M$_{spec}$, and those
derived with the evolutionary tracks, M$_{track}$) can be found in Table \ref{evol_par}. 


\begin{figure}
\epsscale{1.00}
\plotone{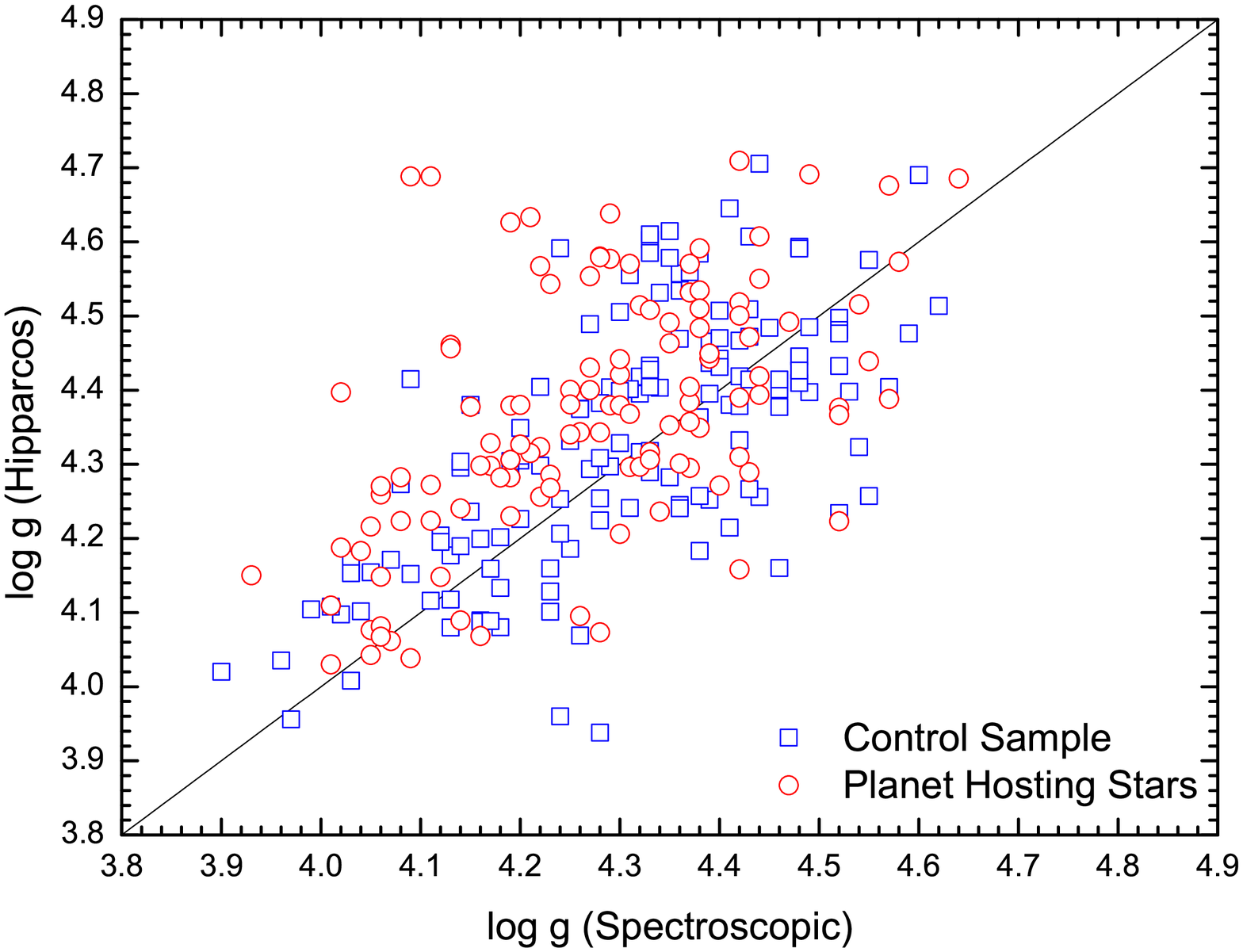}
\caption{A comparison between Hipparcos and spectroscopic gravities for planet hosting stars and control 
sample. The agreement between the two sets of surface gravities is found to be good and systematic differences 
between the two independent scales are less than $\sim$ 0.1 dex on average. The solid line is the bisector
and represents the perfect agreement between the two determinations.}
\label{loggs}
\end{figure}

As the spectroscopic masses have greater errors, we adopt the masses obtained 
with the grid of evolutionary tracks. Using those masses, the Hipparcos surface gravities can be
calculated with the relation:
\begin{equation}
 \label{logg}
 \log g=\log {g_{\sun}}+\log \frac{M}{M_{\sun}}-\log\frac{L}{L_{\sun}}+4\log\frac{T_{\rm eff}}{T_{\rm eff,\sun}},
\end{equation}
where T$_{\rm eff,\sun}$ = 5777 K and $\log g_{\sun}$ = 4.44. The uncertainty in these gravities was calculated 
considering $\sigma(T_{eff,\sun})=0$ and $\sigma(\log g_{\sun})=0$. 
The results are presented in Table \ref{evol_par} (columns 14 and 15). In Figure \ref{loggs} we show a comparison between the
derived spectroscopic log g's (listed in Table \ref{atm_par}) with Hipparcos log g's (listed in Table \ref{evol_par}; column 14). The line indicating
perfect agreement is also shown as a solid line in the figure. The agreement between the two sets of log g's
is good although we note that the Hipparcos gravities are typically found to be higher
(by 0.06 dex on the average) than the spectroscopic values with a standard deviation of $\pm$0.15 dex,
which is of the order of the estimated uncertainties in the derived log g from the iron line analysis.


In addition, masses, radii, ages and trigonometric gravities were also derived with Leo Girardi's web code 
PARAM\footnote{Available at http://stev.oapd.inaf.it/cgi-bin/param}, which is based on a Bayesian 
parameter estimation method (\citealt{dS06}). The mean differences between the results discussed above 
(This work - Girardi's code) are small and indicate good agreement:
$\Delta M = 0.03 \pm 0.05 M_{\sun}$ (N=262), $\Delta R = 0.01 \pm 0.06 R_{\sun}$ (N=223),
$\Delta t = 0.37 \pm 1.46$ Gyr (N=211) and $\Delta \log g = 0.03 \pm 0.05$ (N=262).


\section{Discussion}

\label{disc}

\subsection{Comparisons with Other Studies}

\label{lit}

Several recent studies in the literature 
have derived stellar parameters and metallicities for samples of planet hosting stars. 
In the following we briefly summarize some of these works and then compare their results 
of effective temperatures, surface gravities and metallicities with the
ones obtained in this study. 

{\bf\citet{l03}} determined spectroscopic parameters for 30 stars with giant planets and/or brown dwarf companions. 
Their analysis method is similar to this study, the difference being the line list and \textit{gf}-values 
which were obtained from an inverted solar analysis. 
{\bf \citet{s04,s05}} did a spectroscopic analysis of a large sample of stars with and without planets (119 and 94 
targets, respectively). Their method is very similar to the one used by \citet{l03}, with the difference
being the list of iron lines (the \textit{gf}-values are also solar). 
{\bf \citet{t05}} obtained stellar parameters for a set of 160 mid-F through early-K dwarfs/subgiants.  
The difference with previous studies is the selected iron lines. 
{\bf\citet{vf05}} derived stellar properties for 1040 nearby F, G and K stars observed as part of the Keck Observatory, 
Lick Observatory and Anglo-Australian Telescope (AAT) planet search programs. Their method was different;
stellar parameters and abundances were determined from a direct comparison of observed and synthetic 
spectra across certain spectral intervals using the spectral modelling program, SME. 
In addition, a fixed value of 0.85 km $\rm s^{-1}$ for the microturbulence was adopted for all stars. 
{\bf \citet{lh06}} derived atmospheric parameters for a sample of 216 nearby dwarf stars. They used the standard 
spectroscopic method, but with differential abundances relative to the Sun. 
{\bf \citet{b06}} determined atmospheric parameters from 136 G-type stars from the AAT planet search program. 
In that study, photometric temperatures are obtained from $\bv$ colours 
listed in the \textit{Hipparcos} catalog, with discrete values of microturbulence set at 1.00, 1.25 
and 1.50 km $\rm s^{-1}$. The metallicities and gravities are determined by iterating this parameters until 
the \ion{Fe}{1} and \ion{Fe}{2} abundances are the same.
{\bf \citet{s08}} derived spectroscopic parameters for all 451 solar-type stars from the HARPS Guaranteed 
Time Observations (GTO) \textquotedblleft high-precision\textquotedblright\ sample. Their method closely 
resembles that from \citet{s04,s05}, except for a larger line list and the usage of automatic measurements 
of equivalent widths. 

A direct comparison of the stellar parameters and metallicities obtained for some studied stars in our sample 
with results from other studies discussed above is possible given that there are several targets in common.
Table \ref{comp_lit} shows the average differences (in the sense This study - Literature study) 
computed for the effective temperatures ($\langle\Delta T_{eff}\rangle$), surface gravities ($\langle\Delta \log g\rangle$) and 
metallicities ($\langle\Delta [Fe/H]\rangle$) obtained for all target stars we have in common with 
the studies of \citet{l03}; \citet{s04,s05}; \citet{t05}; \citet{vf05}; \citet{lh06}; \citet{b06}; \citet{s08}; \citet{s04,s05}.
The number of stars compared in each case is found in Table \ref{comp_lit} (column 5). 
Results from this simple and direct comparison are briefly summarized below.

{\bf Effective Temperatures:}
In general, there is not a significant offset in the effective temperature scale
in this study in comparison with the other studies in Table \ref{comp_lit}. 
In particular, for 5 studies we find $\langle\Delta T_{eff}\rangle$ to be less than 15 K, 
which is a quite small 
systematic offset; the \citet{lh06} study has a difference which is only slightly larger ($\sim$35 K). 
A comparison with the  $T_{eff}$ from results in \citet{b06} indicates, however, a more 
significant systematic difference of $\sim$75 K. The 
standard deviations around the mean values are all $\sim$100 K or less, which is in general agreement with
the estimated uncertainties for the derived effective temperatures in this study.

{\bf Surface Gravities:} 
A direct comparison between the average surface gravity value derived for selected targets in this study 
with average results from \citet{l03}; \citet{s04,s05}; \citet{vf05}; \citet{lh06} and \citet{s08}  
indicates that there is a small offset ($\sim$ 0.08 dex -- 0.12 dex) in the $\log g$ scales. 
An agreement (at the level of 0.05 dex or better) is found between our results and \citet{t05} and \citet{b06}.
The stardard deviations of the distributions around the average differences in log g are in all cases less than 0.2 dex; in agreement with the estimated uncertainties in the derived surface gravities.

{\bf Metallicities:} 
In terms of average metallicity values, the iron abundances derived in this study compare very well
with results obtained in other studies in the literature for stars in common. 
There is a slight tendency, however, for the metallicities here to be just marginally lower 
(0.03 dex or less) than the other studies; but such differences are probably statistically insignificant.
Note, however, that the iron abundance results in \citet{b06}  are on average 0.09 dex lower than ours.

\subsection{Metallicity Trends with Effective Temperature and Stellar Mass}

\begin{figure}
\epsscale{1.00}
\plotone{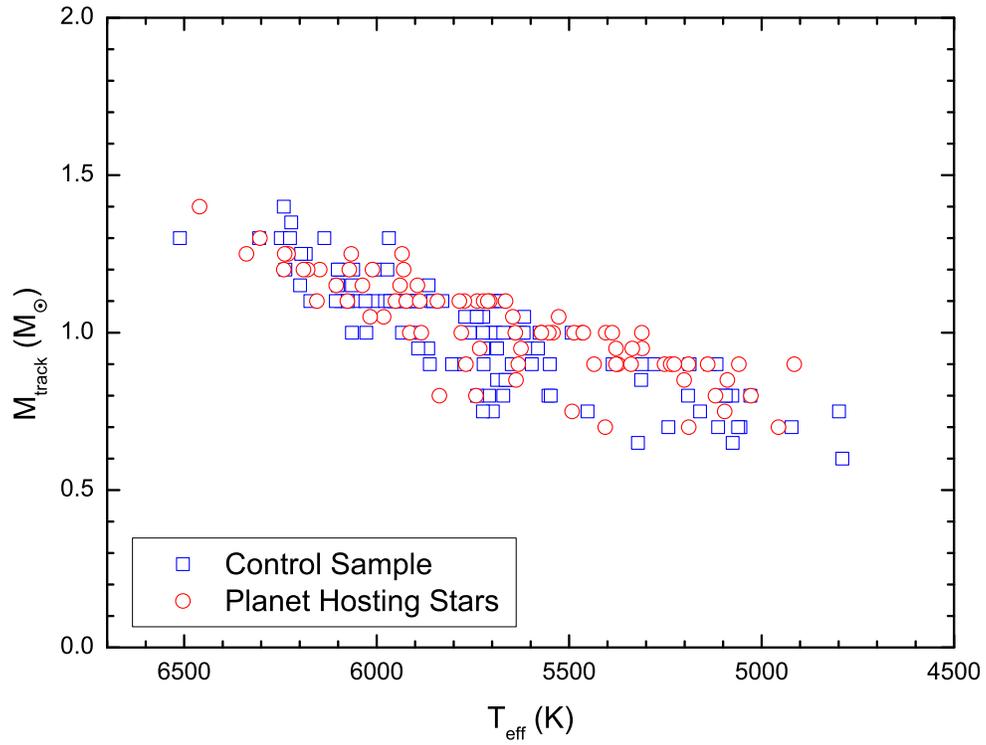}
\caption{The relation between evolutionary track mass and effective temperatures for stars with (red open circles) 
and without (blue open squares) planets. Stars in both samples populate a well-defined 
\textquotedblleft main-sequence\textquotedblright, with significant overlap in mass and T$_{eff}$ between the stars with giant planets
and those without.}
\label{Mtrack_teff}
\end{figure}

Given our sample of 262 stars which have been subjected to a
homogeneous analysis it is possible to search for differences
in the properties of stars with planets compared to those
stars not known to harbor giant planets with periods less than about
4 years.  Two key defining properties of stars are their effective
temperatures and masses, which on the main-sequence are related to
each other, such that increasing T$_{\rm eff}$ maps into
increasing mass, at least over the relatively limited range of
metallicities explored in this sample.  In order to isolate
possible differences between the two samples that might be 
related to T$_{\rm eff}$ or mass, 
stars having surface gravities with $\log g <$ 4.2 were excluded from comparisons in this section.  
The resultant sample consists of 79
stars with planets and 109 stars without planets.  Figure \ref{Mtrack_teff}
compares the properties T$_{\rm eff}$ and derived evolutionary track
mass for the two samples of stars, with mass plotted versus T$_{\rm eff}$.
The main-sequence nature of these stars is obvious from the figure, 
with no significant
differences in the distribution along the main sequence of stars
with and without planets.


\subsubsection{Effective Temperatures, Iron Abundances and Solid-Body Accretion}

\label{teff_feh_accret}

\begin{figure}
\epsscale{0.70}
\plotone{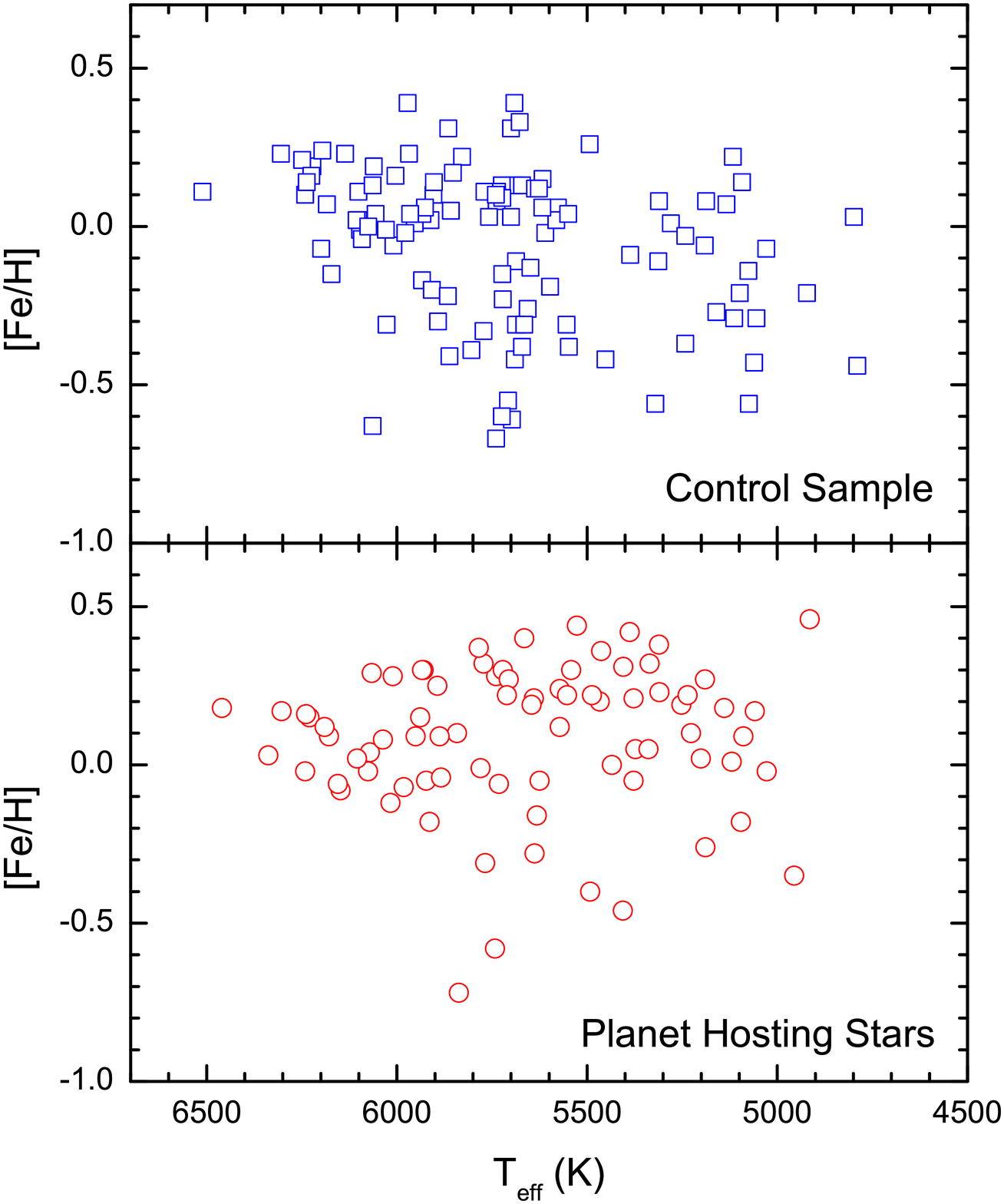}
\caption{The trend between metallicities and effective temperatures for control sample stars (upper panel) and 
planet hosting stars (lower panel). There is a spread of metallicities at each T$_{eff}$, with no strong trends.
Significant solid-body accretion would produce an upturn in the upper envelope of [Fe/H]
with increasing T$_{eff}$: such an effect is not observed in either sample.}
\label{feh_teff}
\end{figure}

The defined set of target stars with log g $\geq$ 4.2 
(or those very near to the main sequence) are now illustrated with their values of [Fe/H] plotted versus
T$_{\rm eff}$ in Figure \ref{feh_teff}. The top panel contains
stars without planets and the bottom panel shows stars with planets.
Since these are main-sequence stars, the effective temperature follows
the stellar mass.  No strong trends between [Fe/H]
and T$_{\rm eff}$ are apparent in either sample, with the lack of an
increase of [Fe/H] with increasing effective temperature placing
limits on the amount of solid-body accretion (material depleted in
H and He, for example) that might have occurred in these stars,
since the convective-zone mass of a main-sequence star is a
strongly decreasing function of increasing T$_{\rm eff}$.  


This same accretion test was conducted by \citet{p01} on an
early small sample ($\sim$30) of stars with planets and they found no
trend of increasing [Fe/H] with T$_{\rm eff}$.  Accretion of solid
material could create a positive trend due to the significantly
decreasing convective zone mass in main-sequence stars with increasing
effective tempratures: e.g., the convective-zone mass decreases by about
a factor of 50 in going from T$_{\rm eff}$ = 5000 K to 6400 K (\citealt{p01}). 
Accretion of only a few Earth masses of solid material
would increase the surface value of [Fe/H] by $\sim$+0.3 dex in a
solar-metallicity star with T$_{\rm eff}$ = 6400 K (see Figure 2 in
\citealt{p01}). No such trend is seen in Figure \ref{feh_teff},
suggesting that accretion of more than a few Earth masses of solid
material is either rare, or such accreted material sinks rapidly
out of the outer convection zone.

\subsubsection{Stellar Mass and Metallicity}

\label{mass_feh}

\begin{figure}
\epsscale{0.70}
\plotone{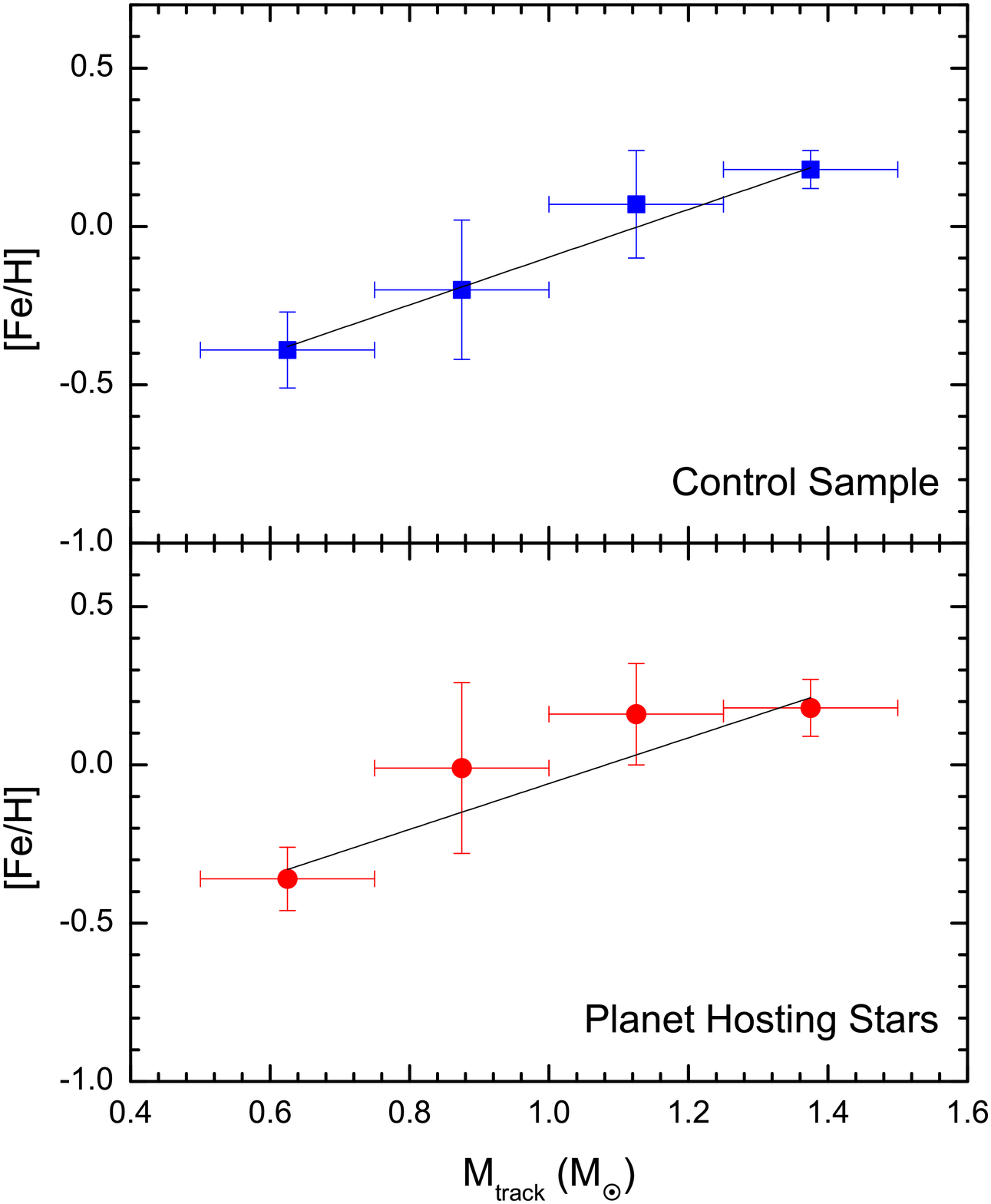}
\caption{Average metallicities versus binned evolutionary track mass for control sample stars 
(upper panel) and planet hosting stars (lower panel). Masses are binned in 0.25 M$_{\sun}$
intervals with each point showing the mean value and standard deviation 
of [Fe/H] within that mass interval. The slopes of $\Delta$[Fe/H]/$\Delta$M are
the same within their uncertainties, in each sample and are likely the result
of the age-metallicity relation. Selective accretion of solid-body material
in stars with giant, closely orbiting planets would result in different
slopes between the two samples and this is not observed.}
\label{feh_avg_Mtrack_bin}
\end{figure}

In addition to the comparison carried out above between [Fe/H]
and T$_{\rm eff}$, it is also instructive to do a similar
comparison with stellar mass (in this case using the evolutionary
track masses); this comparison is shown in Figure \ref{feh_avg_Mtrack_bin}, again, where
stars without planets are plotted in the upper panel and stars
with planets in the lower panel. 
The samples have been binned in mass intervals of 0.25 M$_{\sun}$, as
represented by the error bars in the abscissa. The values
of [Fe/H] plotted represent the mean value within that mass
interval, with the error bars showing the standard deviations
of [Fe/H] at that mass. In both samples, the values of
[Fe/H] increase with increasing stellar mass.  Such an increase
was noted in the review by \citet[his Figure 1]{g06} using
the abundance results from \citet{fv05}.

A signature of solid-body accretion polluting the stellar convective
envelopes would be an upturn in [Fe/H] with increasing stellar
mass, since the convective envelope mass is a rapidly decreasing
function of increasing stellar mass.  At first glance, the increase
in [Fe/H] with mass found here (and noted by \citealt{g06} based 
on the \citealt{fv05} results) might suggest that
solid-body accretion has taken place.  Two effects, however, indicate
that this has not affected the overall metallicities.  First, the
slopes of $\Delta$[Fe/H]/$\Delta$M are identical in both the stars
with planets and stars without planets.  This slope is very roughly
+0.7 dex/M$_{\sun}$ and is similar to the slope that would be
deduced from Figure 1 in \citet{g06}. The fact that all of the
various samples of stars, with and without giant planets, exhibit similar
behavior in metallicity with mass argues that pollution has not 
selectively altered the values
of [Fe/H] in a significant way for the stars with giant planets.  The
second point to note is that a positive slope of [Fe/H] with stellar
mass would result from an age-metallicity relation.  Since more massive
stars have shorter main-sequence lifetimes, they would be biased towards
higher values of [Fe/H], while lower-mass stars would be a mixture of old
and young stars, which would shift the overall distribution to lower
average values of [Fe/H].

In summary, comparisons among iron abundances with effective temperatures
and stellar masses in both samples of stars (with and without large
planets, respectively) reveal that accretion of solid-body material does
not affect significantly ($\sim$0.1-0.2 dex) the overall bulk metallicity
in either sample.  This does not rule out smaller amounts of accretion,
which might affect abundance ratios between certain types of elements (such
as volatile versus refractory species, as suggested by \citealt{s01}; see \citealt{m09}
for an alternative interpretation). 
This question will be addressed in a later paper
using the spectra from this dataset and analyzing a broad range of elements.


%
\subsection{Metallicity Distributions of Sample Stars}

\label{met}

\begin{figure}
\epsscale{1.00}
\plotone{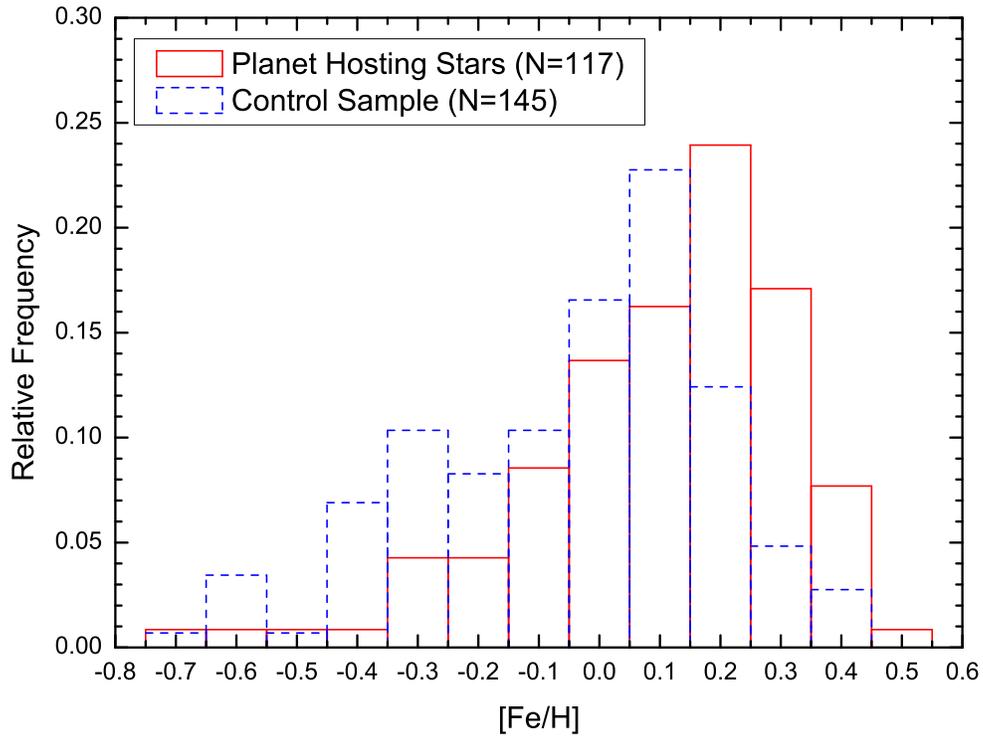}
\caption{Metallicity distributions obtained for planet hosting stars (red solid line) and a control sample
of disk stars not known to host giant planets (blue dashed line). The peaks of the metallicity distributions
are offset by 0.15 dex indicating that the sample of stars which host planets is typically more metal rich.}
\label{feh_hist}
\end{figure}

As discussed in Section \ref{lit} the metallicities ([Fe/H]) derived here are generally consistent 
(within the expected errors) with metallicities found in other studies of planet hosting
stars in the literature.
As the present study relies on a homogeneous and self-consistent analysis of a sample of 262 stars, 
having comparable numbers of planet hosting and comparison disk stars, it is possible
to quantify differences in the metallicity distributions in these two populations.
Figure \ref{feh_hist} shows the metallicity distributions 
for stars with planets (solid line histogram) and comparison stars (dashed line histogram). 
There is an offset in the peak metallicity of the two histograms in the figure.
The peak of the distrbution for stars with planets is located in 
the bin centered at [Fe/H] = +0.20 dex and the average metallicity of this sample is $\langle[Fe/H]\rangle$ = +0.11 dex. 
For the comparison stars, the peak is on the bin centered at [Fe/H] = +0.10 dex and the average metallicity 
in this case is lower: $\langle[Fe/H]\rangle = -0.04$ dex. 
Thus, there is an offset of 0.15 dex between the mean metallicities of the two samples.

When comparing properties, such as metallicity, between samples of planet hosting stars and those without
giant planets, it is worth noting some selection biases inherent in these samples. As summarized by
\citet{g06}, Doppler surveys avoid young, chromospherically active stars (which also
typically are fast rotators) and contain only small numbers of metal poor stars (with [Fe/H] $<$ -0.5 dex) 
because these objects are rare in the solar neighbourhood. 
In particular, our control sample of disk stars would also suffer from such biases since it was selected 
in order to search for the presence of planets.


As this offset between the peak values of the two histograms is of the order of, or just slightly
higher than, the expected uncertainties in the derived iron abundances  themselves,
it is useful to perform a robust statistical test in order to further investigate whether the 
metallicity offset is meaningful. In this sense, we conducted a Two-Sample 
Kolmogorov-Smirnov test and found a probability $P = 6.17 \times 10^{-6}$ that the two samples are drawn from the 
same parent population. This low probability confirms the results previously found in the literature that the
population of stars hosting giant planets is more metal-rich than the population of stars not known to harbour such 
planets. 

If a volume limited sample with a radius of 18 pc is defined here for comparison, the
mean metallicity for the control sample disk stars (N=46) is now -0.11 dex and the offset relative to 
planet-hosting stars becomes 0.22 dex, similar to the 
one found by \citet{fv05} based on a much larger sample. The average metallicity 
found for the volume-limited sample in this study is also consistent with the results from \citet{s04,s05} and \citet{s08}. 
The former study uses a comparison sample of 94 stars within 20 pc of the Sun and finds $\langle[Fe/H]\rangle 
= -0.11$ dex. The latter work extends the comparison sample to 385 stars and the enclosed radius to 56 pc and finds 
$\langle[Fe/H]\rangle = -0.12$ dex.

The results from Sections \ref{teff_feh_accret} and \ref{mass_feh} indicate that solid body accretion has probably not altered surface values of [Fe/H] at the level of the offset in metallicity; the difference in [Fe/H] between the two samples suggests that intrinsic metallicity influences giant planet formation and migration.

\subsubsection{Metallicities of stars hosting Neptunian-mass planets}

\label{nept_feh}

The conclusions from the previous section favor the premise that metallicity
plays a role in influencing the formation of giant planets, i.e., planets with roughly the mass of Jupiter. 
Within this model it is worthwhile exploring whether stellar metallicity also plays a role in the 
mass distribution of planetary systems. Such a comparison begins to probe how 
the underlying planetary system architecture might depend on metallicity. In this section, the metallicities
of stars harboring only lower mass planets, i.e. those having Neptunian masses, 
are compared to systems containing the larger Jovian mass planets. 

\citet{s08} presented preliminary results in which they found possible metallicity differences between 
stars hosting Jovian mass planets compared to those hosting Neptunian-mass planets ($M_{p}\sin i< 25M_{\earth}$). 
The differences between the two metallicity distributions defined respectively by stars with Jovian-mass planets as opposed to 
Neptunian-mass planets indicated that these groups are not likely to belong to the same populations of stars. 
This conclusion, however, was based on a comparison of  63 Jovian hosting stars with a sample of 11 Neptunian hosting 
stars (those which contained at least one Neptune mass planet). Five of these 11 stars were M dwarfs and their
metallicities were taken from the literature, while three were FG dwarfs analyzed by \citet[HARPS GTO]{s08}. 
The difference between the two metallicitiy distributions was a small mean offset of about 0.1 dex,
with the Neptune hosting stars having a slightly lower mean metallicity.

Because M dwarfs are more difficult to analyze spectroscopically, due to considerable line blending and 
blanketing from molecules, it is necessary to consider abundance uncertainties and 
systematics when using results from these complex stellar spectra. Recent abundance analyses of M dwarfs
include \citet{b05}; \citet{ww06}; \citet{bean06} and \citet{ja09}. The studies by \citet{b05} and \citet{ja09} point, for instance, to potentially
large uncertainties in derived M dwarf abundances. For example, \citet{ja09} 
find an average offset of $\sim$ 0.30 dex in [Fe/H] when their abundances are compared to the same M dwarfs from
\citet[see Table \ref{nept_table}]{b05}. Such discrepancies suggest that, until results for the cooler M dwarfs 
are on firmer ground, it is prudent to investigate the effects of both including M dwarf  metallicities in 
such comparisons, as well as excluding them. 

The sample studied here contains 9 systems which host at least one Neptunian mass planet, none
of which are M dwarfs.
This is the largest sample of stars hosting Neptunian size planets analyzed homogeneously for metallicities to date
and this subsample can be directly compared to the Jupiter-like planet hosting stellar sample. The
strength of such a comparison rests upon the fact that all stars have been analyzed homogeneously
and are within a similar range of stellar parameters, so that systematic errors are not likely to produce
spurious differences in the metallicity distributions. The weakness is that the sample of Neptune-mass
hosts has only a small number of stars.

\begin{figure}
\epsscale{0.70}
\plotone{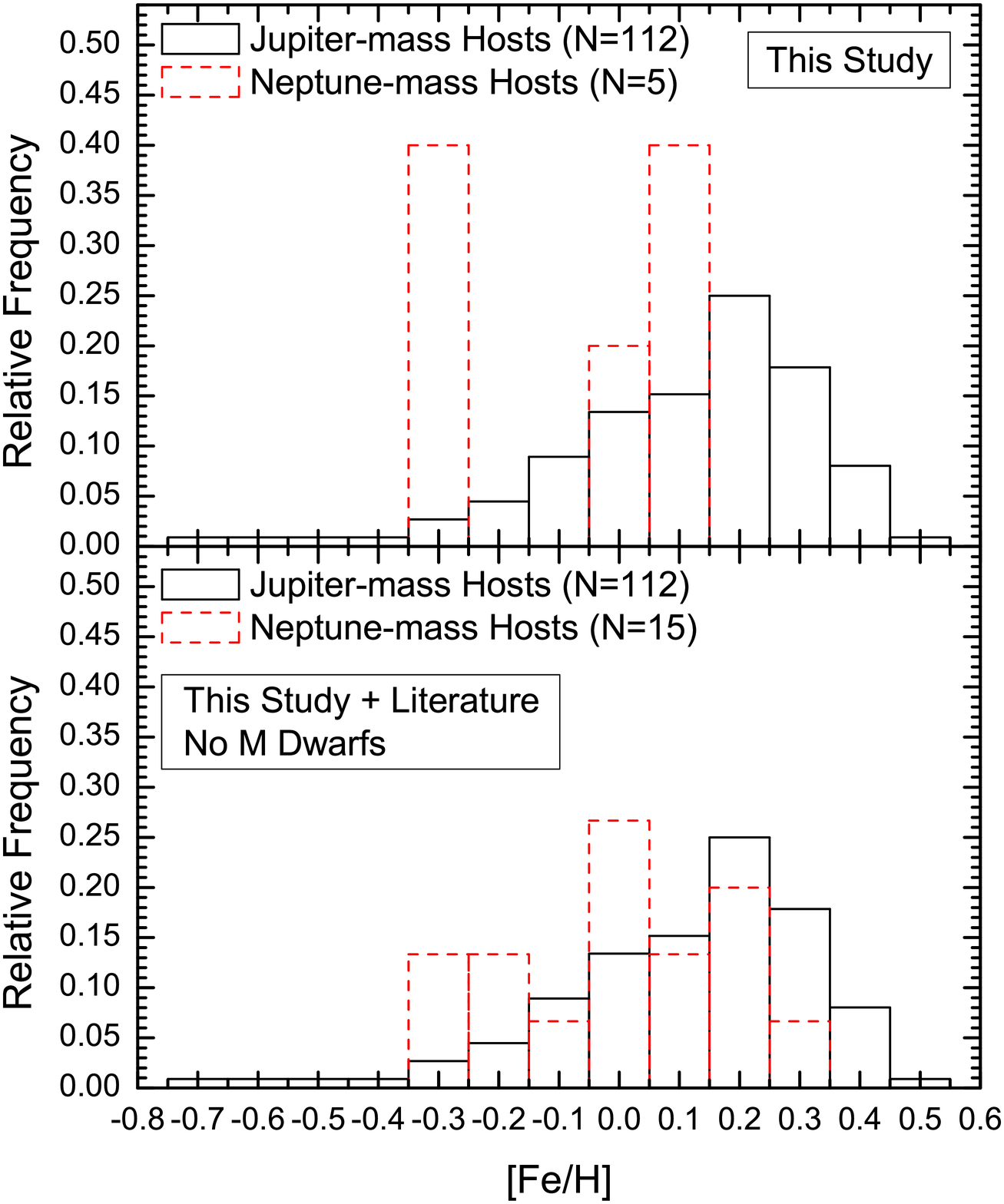}
\caption{A comparison between the metallicity distributions for planet hosting stars.
{\bf Top panel}: The black solid line histogram represents the sample of planet hosting stars containing at least 
one Jupiter planet. The red dashed line represents the sample of stars hosting {\bf only} planets with Neptune-like masses. 
The metallicities are from this study. {\bf Bottom panel}: Results for metallicities in Neptunian-mass hosting
stars from the literature are added (red dashed line histogram). Given the large uncretainties in the
M dwarf metallicities these stars have not been added to the sample. 
There is a hint that there is offset between the two distributions represented both in the top and bottom panels.}
\label{nept_hist}
\end{figure}

Figure \ref{nept_hist} (top panel) shows histograms representing metallicity distributions of two samples: 
those stars hosting at least one Jupiter-mass planet (N=112; black solid line) 
and stars hosting only Neptune-mass planets (N=5; represented by the dashed red line).
There is a hint that stars with only Neptunian planets tend to be more metal poor compared to 
Jovian-planet hosting stars. The average metallicity of the Neptunian hosts is -0.08 dex, while
the Jupiter host metallicity distribution has an average of +0.12 dex.  
If a Two-Sample Kolmogorov-Smirnov test is performed, we find a probability of 8\% that the 
stars in the two samples belong to the same metallicity population (which agrees with \citealt{s08}). 
This is a tantalizing result that suggests that metallicity may play a role not just in the formation
of giant planets, but may also influence the distribution of planetary masses within exo-solar systems. 
This important question needs to be answered more definitively, but this will require larger samples.


In order to extend the sample of stars with Neptunian mass planets (shown in the top panel of Figure \ref{nept_hist}) 
in this discussion, a list of stars with at least one Neptunian planet was compiled from The Extrasolar Planet 
Encyclopaedia and is presented in Table \ref{nept_table} along with the metallicity results from the
different studies in the literature.  In a similar analysis as was done for the sample of stars studied here
(discussed above), Two-Sample Kolmogorov-Smirnov tests were done now with the inclusion of the literature sample 
using different permutations.  The results from these tests are discussed below. 

1) Using the literature values of [Fe/H] for only F, G, and K dwarfs (no M dwarfs) 
a difference in the mean [Fe/H] of +0.11 dex (in the sense of
Jovian-mass hosts minus Neptunian-mass hosts) is found, with a probability of P=17\% that
the two samples were drawn from the same [Fe/H] populations (with N(Jovian-mass)=112 and
N(Neptunian-mass)=15). The histogram showing the comparison of these two distributions
is presented in the bottom panel of \ref{nept_hist}. 

2) Using all literature values, with M dwarf abundances from \citet{ja09} included, 
we find $\Delta$[Fe/H]=+0.10 dex and P=17\% (N(Jovian-mass=112 and
N(Neptunian-mass)=19). 

3) When M-dwarf abundances from \citet{b05,b07} are used instead,\linebreak
$\Delta$[Fe/H]=+0.14 dex and P=5\% (N(Jovian-mass)=112 and N(Neptunian-mass)=18). 

In the above exercise the stars with planets were divided into systems with Jovian-mass
planets and those with Neptunian-mass planets, respectively. The metallicity
comparison can also be carried out by dividing the sample into stars with at least
one Neptunian-mass planet, regardless of whether there is also a Jovian mass planet
in the system, and those systems with Jovian mass planets but no Neptunian-mass planets.
The average metallicity of the sample of stars which host at least one Neptunian planet 
(N=9) is +0.09 dex,  close to the value derived for stars hosting at least one Jupiter-mass planet 
(+0.12 dex; N=112).
A Two-Sample Kolmogorov-Smirnov test reveals that the probability that these two samples 
belong to the same parent population is 91\%.  This comparison strengthens the idea
that metallicity played a role in setting the mass of the most massive planet in a system.

All of these various comparisons taken together suggest that lower values of [Fe/H] tend to 
produce lower-masses of the most massive planet within a planetary system. Such conclusions, however,
should be viewed with caution since the number of stars harboring Neptune-size planets
analyzed to date is still rather small, but in line with what would be
expected from models of planet formation via core accretion 
(\citealt{il04}; \citealt{mordasini09a,mordasini09b}; see also review by \citealt{boss10}).


\section{Conclusions}

\label{conc}

We have determined stellar parameters for 117 main-sequence stars with
planets discovered via radial velocity surveys and 145
comparison stars which have been found to exhibit nearly
constant radial velocities and are not likely to host
large, closely orbiting planets. The stellar parameters
were derived from a classical spectroscopic analysis,
using accurate laboratory \textit{gf}-values of Fe lines and
automatically measured equivalent widths, after
critical evaluations of their quality.  The values of 
effective temperature, surface gravity (as $\log g$), 
microturbulent velocities, and iron abundances are in
general good agreement with most of the values presented
in a number of literature studies, but problems with a 
few individual stars may remain. 

Correlations between [Fe/H] and T$_{\rm eff}$ in 
members of either sample are not found,
which places stringent limits on the possible accretion
of solid material (of less than a few Earth masses)
onto the surfaces of these stars.  
A trend of increasing [Fe/H] with increasing stellar mass
is found in both samples of stars, with the slope of $\Delta$[Fe/H]/$\Delta$M
being the same for stars with giant planets and control sample
stars. The same value of $\Delta$[Fe/H]/$\Delta$M for
both samples rules out solid-body accretion, which leaves the
underlying disk age-metallicity relation as the likely cause
of the positive correlation of [Fe/H] with mass.

It should be noted that the samples analyzed here
were not selected based on any rigourous criteria, other
than being segregated based on the presence of giant planets
in one sample and the probable absence of such planets in the
other.  The list of stars without planets preferentially
included metal-rich stars, while some stars with planets were
discovered by surveys that were based on high-metallicity as
a criterion.

A comparison of iron abundances between the stars with 
planets with the results from the control sample of disk stars 
confirms the results obtained by previous studies showing that planet-hosting
stars exhibit larger metallicities than stars not
harboring planets; the difference found in this sample
is that stars with planets are shifted by +0.15 dex in
[Fe/H] when compared to the stars without planets.

The sample of stars with planets discussed here contains 9 stars
which host at least one Neptunian-mass planet; of these 9 systems,
4 also contain a Jovian-mass planet, while 5 contain only Neptunian-sized 
planets and no Jovian-mass closely orbiting planets.  A
statistical test of the iron abundances indicates that there is probably
a real difference between the metallicity distributions of stars
which contain {\it only} smaller Neptunian-sized planets in comparison with
stars hosting the larger Jovian-mass planets (see also \citealt{s08}). 

Although it should be recognized that the sample sizes are
still small, there seems to be early indications that metallicity
plays an important role in setting the mass of the most massive planet.
It is also important to note that such a
conclusion obtained here is based on stars which have all been analyzed
homogeneously and which also have similar stellar parameters, therefore
avoiding the large uncertainties still hampering the analyses and derived
metallicities for the cooler M dwarfs.  The same statistical test
applied to metallicity results in the literature obtained for all FGK stars 
which host only Neptunian-sized planets allows for a larger sample and 
overall corroborates the results obtained with our sample.


\acknowledgements{We would like to dedicate this paper to the co-author 
Chico Ara\'ujo who was our friend, adviser and colleague and passed away in 2009. 
We acknowledge the financial support of CNPq. Research presented
here was supported in-part by NASA grant NNH08AJ581. Luan Ghezzi
thanks Oliver Sch\"{u}tz for his valuable help in
using the FEROS DRS package, and Cl\'audio Bastos for conducting
the observing run in February 2008. We thank the anonymous referee for useful comments 
that helped improving the paper.}


\begin{deluxetable}{lcccc}
\tablecolumns{5}
\tablewidth{0pt}
\tablecaption{Log of Observations\label{obslog}}
\tablehead{
\colhead{Star} & \colhead{$V$} & \colhead{Observation} & \colhead{$T_{exp}$} & \colhead{S/N} \\
\colhead{} & \colhead{} & \colhead{Date} & \colhead{(s)} & \colhead{($\sim$ 6700 \AA)}}
\startdata
\sidehead{\textit{Planet Hosting Stars}}
HD 142       &  5.70 & 2007 Aug 30 &   200 & 367 \\
HD 1237      &  6.59 & 2007 Aug 30 &   480 & 466 \\
HD 2039      &  9.00 & 2007 Aug 28 &  3000 & 362 \\
HD 2638      &  9.44 & 2007 Aug 29 &  3000 & 276 \\
HD 3651      &  5.88 & 2007 Aug 29 &   200 & 356 \\
\sidehead{\textit{Control Sample}}
HD 1581      &  4.23 & 2008 Aug 20 &    30 & 216 \\
HD 1835      &  6.39 & 2007 Oct 02 &   200 & 388 \\
HD 3823      &  5.89 & 2007 Oct 02 &   200 & 452 \\
HD 4628      &  5.74 & 2008 Aug 20 &   200 & 375 \\
HD 7199      &  8.06 & 2008 Aug 19 &  1200 & 348 \\
\enddata
\tablecomments{Table \ref{obslog} is published in its entirety in the eletronic edition of the Astrophysical 
               Journal. A portion is show here for guidance regarding its form and content.}
\end{deluxetable}


\begin{deluxetable}{ccccc}
\tablecolumns{5}
\tablewidth{0pt}
\tablecaption{Selected Fe lines and Measured Equivalent Widths for the Sun.\label{linelist}}
\tablehead{
\colhead{$\lambda$} & \colhead{Ion} & \colhead{LEP} & \colhead{$\log gf$} & \colhead{W$_{\lambda}\sun$} \\
\colhead{(\AA)} & \colhead{} & \colhead{(eV)} & \colhead{(dex)} & \colhead{(m\AA)}} 
\startdata
4779.439 & \ion{Fe}{1} & 3.415 & $-$2.020 & 40.4 \\
4788.751 & \ion{Fe}{1} & 3.237 & $-$1.763 & 63.3 \\
4802.875 & \ion{Fe}{1} & 3.695 & $-$1.514 & 58.3 \\
4962.572 & \ion{Fe}{1} & 4.178 & $-$1.182 & 52.7 \\
5054.642 & \ion{Fe}{1} & 3.640 & $-$1.921 & 39.6 \\
5247.049 & \ion{Fe}{1} & 0.087 & $-$4.946 & 65.8 \\
5379.574 & \ion{Fe}{1} & 3.694 & $-$1.514 & 59.8 \\
5618.631 & \ion{Fe}{1} & 4.209 & $-$1.276 & 48.4 \\
5741.846 & \ion{Fe}{1} & 4.256 & $-$1.670 & 30.8 \\
5775.081 & \ion{Fe}{1} & 4.220 & $-$1.298 & 57.9 \\
5778.453 & \ion{Fe}{1} & 2.588 & $-$3.430 & 21.3 \\
5855.076 & \ion{Fe}{1} & 4.608 & $-$1.478 & 21.4 \\
5916.247 & \ion{Fe}{1} & 2.453 & $-$2.994 & 55.0 \\
5956.692 & \ion{Fe}{1} & 0.859 & $-$4.605 & 51.3 \\
6120.246 & \ion{Fe}{1} & 0.915 & $-$5.970 &  5.2 \\
6151.617 & \ion{Fe}{1} & 2.176 & $-$3.299 & 48.7 \\
6173.334 & \ion{Fe}{1} & 2.223 & $-$2.880 & 67.2 \\
6200.313 & \ion{Fe}{1} & 2.608 & $-$2.437 & 72.5 \\
6219.279 & \ion{Fe}{1} & 2.198 & $-$2.433 & 89.8 \\
6240.645 & \ion{Fe}{1} & 2.223 & $-$3.170 & 48.0 \\
6265.131 & \ion{Fe}{1} & 2.176 & $-$2.550 & 84.2 \\
6380.743 & \ion{Fe}{1} & 4.186 & $-$1.376 & 51.8 \\
6593.870 & \ion{Fe}{1} & 2.433 & $-$2.422 & 84.5 \\
6699.141 & \ion{Fe}{1} & 4.593 & $-$2.101 &  7.7 \\
6739.520 & \ion{Fe}{1} & 1.557 & $-$4.794 & 11.7 \\
6750.150 & \ion{Fe}{1} & 2.424 & $-$2.621 & 73.2 \\
6858.145 & \ion{Fe}{1} & 4.607 & $-$0.930 & 51.1 \\
4993.358 & \ion{Fe}{2} & 2.807 & $-$3.670 & 37.6 \\
5132.669 & \ion{Fe}{2} & 2.807 & $-$4.000 & 26.9 \\
5284.109 & \ion{Fe}{2} & 2.891 & $-$3.010 & 62.6 \\
5325.553 & \ion{Fe}{2} & 3.221 & $-$3.170 & 39.2 \\
5414.073 & \ion{Fe}{2} & 3.221 & $-$3.620 & 27.6 \\
5425.257 & \ion{Fe}{2} & 3.199 & $-$3.210 & 41.5 \\
5991.376 & \ion{Fe}{2} & 3.153 & $-$3.560 & 30.4 \\
6084.111 & \ion{Fe}{2} & 3.199 & $-$3.800 & 20.5 \\
6149.258 & \ion{Fe}{2} & 3.889 & $-$2.720 & 35.0 \\
6369.462 & \ion{Fe}{2} & 2.891 & $-$4.190 & 19.9 \\
6416.919 & \ion{Fe}{2} & 3.892 & $-$2.680 & 38.5 \\
6432.680 & \ion{Fe}{2} & 2.891 & $-$3.580 & 40.4 \\
\enddata
\end{deluxetable}


\begin{deluxetable}{lccccccccr}
\tablecolumns{10}
\rotate
\tablewidth{0pt}
\tablecaption{Atmospheric Parameters.\label{atm_par}}
\tablehead{
\colhead{Star} & \colhead{T$_{eff}$} & \colhead{$\log$ g} & \colhead{$\xi$} & \colhead{A(Fe)} & \colhead{$\sigma$} & \colhead{N} & \colhead{$\sigma$} & \colhead{N} & \colhead{[Fe/H]} \\
\colhead{} & \colhead{(K)} & \colhead{} & \colhead{(km $\rm s^{-1}$)} & \colhead{} & \colhead{(\ion{Fe}{1})} & \colhead{(\ion{Fe}{1})} & \colhead{(\ion{Fe}{2})} & \colhead{(\ion{Fe}{2})} & \colhead{}}
\startdata
\sidehead{\textit{Planet Hosting Stars}}
HD 142  & 6338 $\pm$  46 & 4.34 $\pm$ 0.14 & 2.27 $\pm$ 0.08 & 7.46 & 0.09 & 23 & 0.07 & 10 &    0.03 $\pm$ 0.04 \\
HD 1237 & 5572 $\pm$  40 & 4.58 $\pm$ 0.09 & 1.34 $\pm$ 0.04 & 7.55 & 0.09 & 27 & 0.05 &  9 &    0.12 $\pm$ 0.04 \\
HD 2039 & 5934 $\pm$  36 & 4.30 $\pm$ 0.13 & 1.26 $\pm$ 0.04 & 7.73 & 0.08 & 27 & 0.05 & 12 &    0.30 $\pm$ 0.03 \\
HD 2638 & 5236 $\pm$  70 & 4.38 $\pm$ 0.19 & 0.86 $\pm$ 0.04 & 7.65 & 0.07 & 26 & 0.07 &  9 &    0.22 $\pm$ 0.05 \\
HD 3651 & 5252 $\pm$  65 & 4.32 $\pm$ 0.18 & 0.81 $\pm$ 0.02 & 7.62 & 0.08 & 27 & 0.05 & 10 &    0.19 $\pm$ 0.03 \\
\sidehead{\textit{Control Sample}}
HD 1581 & 5908 $\pm$  31 & 4.26 $\pm$ 0.13 & 1.17 $\pm$ 0.04 & 7.23 & 0.08 & 25 & 0.08 & 12 & $-$0.20 $\pm$ 0.03 \\
HD 1835 & 5829 $\pm$  41 & 4.39 $\pm$ 0.16 & 1.24 $\pm$ 0.04 & 7.65 & 0.07 & 22 & 0.05 & 10 &    0.22 $\pm$ 0.03 \\
HD 3823 & 6012 $\pm$  31 & 4.18 $\pm$ 0.08 & 1.92 $\pm$ 0.10 & 7.08 & 0.07 & 25 & 0.05 & 11 & $-$0.35 $\pm$ 0.02 \\
HD 4628 & 5055 $\pm$  40 & 4.33 $\pm$ 0.19 & 0.88 $\pm$ 0.04 & 7.14 & 0.08 & 25 & 0.06 &  5 & $-$0.29 $\pm$ 0.02 \\
HD 7199 & 5349 $\pm$  65 & 4.09 $\pm$ 0.19 & 1.04 $\pm$ 0.04 & 7.74 & 0.09 & 26 & 0.05 & 10 &    0.31 $\pm$ 0.05 \\
\enddata
\tablecomments{Table \ref{atm_par} is published in its entirety in the eletronic edition of the Astrophysical 
               Journal. A portion is show here for guidance regarding its form and content.}
\end{deluxetable}


\begin{deluxetable}{lcccrrcccccccccrc}
\tabletypesize{\scriptsize}
\tablecolumns{17}
\rotate
\tablewidth{0pt}
\tablecaption{Evolutionary Parameters.\label{evol_par}}
\tablehead{
\colhead{Star} & \colhead{$\pi$} & \colhead{$\sigma_{\pi}$} & \colhead{$A_{V}$} & \colhead{$BC_{V}$} & \colhead{$\log(L/L_{\sun})$} & \colhead{$\sigma_{\log(L/L_{\sun})}$} & \colhead{$R$} & \colhead{$\sigma_{R}$} & \colhead{$M_{spec}$} & \colhead{$\sigma(M_{spec})$} & \colhead{M$_{track}$} & \colhead{$\sigma(M_{track})$} & \colhead{$\log g_{Hipp}$} & \colhead{$\sigma(\log g_{Hipp})$} & \colhead{Age} & \colhead{$\Delta$Age} \\
\colhead{} & \colhead{(mas)} & \colhead{(mas)} & \colhead{} & \colhead{} & \colhead{} & \colhead{} & \colhead{(R$_{\sun}$)} & \colhead{(R$_{\sun}$)} & \colhead{(M$_{\sun}$)} & \colhead{(M$_{\sun}$)} & \colhead{(M$_{\sun}$)} & \colhead{(M$_{\sun}$)} & \colhead{} & \colhead{} & \colhead{(Gyr)} & \colhead{(Gyr)}}
\startdata
\sidehead{\textit{Planet Hosting Stars}}
HD 142  &  38.89 & 0.37 & 0.05 &    0.017 &    0.462 & 0.061 & 1.41 & 0.11 & 1.59 & 0.77 & 1.25 & 0.10 & 4.24 & 0.08 &     2.5 & 2.0-3.5  \\
HD 1237 &  57.15 & 0.31 & 0.04 & $-$0.075 & $-$0.196 & 0.060 & 0.86 & 0.07 & 1.02 & 0.49 & 1.00 & 0.10 & 4.57 & 0.08 & $<$ 1.0 & 0.0-3.0  \\
HD 2039 &   9.75 & 0.95 & 0.11 & $-$0.004 &    0.376 & 0.104 & 1.46 & 0.18 & 1.55 & 0.81 & 1.25 & 0.10 & 4.21 & 0.11 &     3.5 & 3.0-4.5  \\
HD 2638 &  20.03 & 1.49 & 0.10 & $-$0.159 & $-$0.368 & 0.089 & 0.80 & 0.09 & 0.55 & 0.28 & 0.90 & 0.10 & 4.59 & 0.11 & $<$ 1.0 & 0.0-4.0  \\
HD 3651 &  90.42 & 0.32 & 0.02 & $-$0.153 & $-$0.287 & 0.060 & 0.87 & 0.07 & 0.57 & 0.28 & 0.90 & 0.10 & 4.51 & 0.08 &     5.0 & 0.0-12.0 \\
\sidehead{\textit{Control Sample}}
HD 1581 & 116.46 & 0.16 & 0.01 & $-$0.037 &    0.102 & 0.060 & 1.08 & 0.08 & 0.76 & 0.37 & 1.00 & 0.10 & 4.38 & 0.08 &     6.5 & 4.0-9.0  \\
HD 1835 &  47.93 & 0.53 & 0.08 & $-$0.025 &    0.033 & 0.061 & 1.02 & 0.08 & 0.93 & 0.45 & 1.10 & 0.10 & 4.47 & 0.08 &     1.0 & 0.0-3.5  \\
HD 3823 &  40.07 & 0.34 & 0.04 & $-$0.035 &    0.376 & 0.061 & 1.42 & 0.11 & 1.11 & 0.54 & 1.00 & 0.10 & 4.13 & 0.08 &     7.5 & 6.5-9.0  \\
HD 4628 & 134.14 & 0.51 & 0.01 & $-$0.226 & $-$0.549 & 0.060 & 0.69 & 0.06 & 0.37 & 0.18 & 0.70 & 0.10 & 4.60 & 0.09 &     9.0 & 0.0-14.0 \\
HD 7199 &  28.33 & 0.57 & 0.10 & $-$0.122 & $-$0.132 & 0.063 & 1.00 & 0.08 & 0.45 & 0.22 & 0.95 & 0.10 & 4.41 & 0.08 &     9.0 & 5.0-13.0 \\
\enddata
\tablecomments{Table \ref{evol_par} is published in its entirety in the eletronic edition of the Astrophysical 
               Journal. A portion is show here for guidance regarding its form and content.}
\end{deluxetable}


\begin{deluxetable}{lcccc}
\tablecolumns{5}
\tablewidth{0pt}
\tablecaption{Comparison with other results in the literature.\label{comp_lit}}
\tablehead{
\colhead{Study} & \colhead{$\langle\Delta T_{eff}\rangle$ (K)} & \colhead{$\langle\Delta\log g\rangle$} & \colhead{$\langle\Delta$[Fe/H]$\rangle$} & \colhead{N$_{Stars}$}}
\startdata
\citet{l03}      & $ -5\pm$74 & $-$0.10$\pm$0.15 & $-$0.03$\pm$0.05 &  23      \\
\citet{s04,s05}  &  $-2\pm$72 & $-$0.08$\pm$0.13 & $-$0.02$\pm$0.06 & 113      \\
\citet{t05}      &  $-8\pm$65 & $-$0.05$\pm$0.16 & $-$0.03$\pm$0.07 &  35      \\
\citet{vf05}     &  10$\pm$65 & $-$0.11$\pm$0.14 & $-$0.01$\pm$0.06 & 223      \\
\citet{lh06}     & $-32\pm$84 & $-$0.11$\pm$0.15 & $-$0.02$\pm$0.07 &  56      \\
\citet{b06}      & 74$\pm$113 &    0.01$\pm$0.19 &    0.09$\pm$0.09 &  90      \\
\citet{s08}      & $-14\pm$61 & $-$0.11$\pm$0.11 &    0.00$\pm$0.06 & 119      \\
\enddata
\tablecomments{$\Delta$ = This study - Literature Study}
\end{deluxetable}


\begin{deluxetable}{lccrl}
\tablecolumns{5}
\tablewidth{0pt}
\tablecaption{Neptunian-mass Planet Hosts\label{nept_table}}
\tablehead{
\colhead{Star} & \colhead{$M_{P}\sin \textit{i}$} & \colhead{Jupiter} & \colhead{[Fe/H]} & \colhead{Reference} \\
\colhead{} & \colhead{($M_{\earth}$)} & \colhead{} & \colhead{} & \colhead{[Fe/H]} }
\startdata
\sidehead{\textit{Results from This Work}}
HD 4308      & 12.87 &  no & $-$0.31 &                         \\
HD 16417     & 21.93 &  no &    0.14 &                         \\
HD 40307     &  4.20 &  no & $-$0.35 &                         \\
HD 47186     & 22.78 & yes &    0.21 &                         \\
HD 69830     & 10.49 &  no &    0.00 &                         \\
HD 125612    & 21.29 & yes &    0.25 &                         \\
HD 160691    & 10.56 & yes &    0.23 &                         \\
HD 181433    &  7.56 & yes &    0.46 &                         \\
HD 219828    & 20.98 &  no &    0.14 &                         \\
\sidehead{\textit{Literature Results}}
HD 7924      &  9.22 &  no & $-0.15$ & \citet{h09}             \\
HD 1461      &  7.60 &  no &    0.18 & \citet{vf05}            \\
             &       &     &    0.21 & \citet{lh06}            \\
             &       &     &    0.19 & \citet{s08}             \\
             &       &     &    0.19 & Average                 \\
CoRoT-7      &  4.80 &  no &    0.05 & \citet{l09}             \\
55 Cnc       &  7.63 & yes &    0.33 & \citet{s04}             \\
BD-082823    & 14.30 & yes & $-$0.07 & \citet{h10}             \\
HD 90156     & 17.48 &  no & $-$0.24 & Encyclopaedia           \\
61 Vir       &  5.09 &  no &    0.01 & \citet{s04,s05}         \\
             &       &     &    0.05 & \citet{t05}             \\
             &       &     &    0.11 & \citet{vf05}            \\
             &       &     & $-$0.02 & \citet{s08}             \\
             &       &     &    0.04 & Average                 \\
HD 125595    & 14.30 &  no &    0.02 & Encyclopaedia           \\
HD 156668    &  4.16 &  no & $-$0.07 & \citet{m08}             \\
Kepler-4     & 24.47 &  no &    0.17 & \citet{b10}             \\
HD 179079    & 25.43 &  no &    0.25 & \citet{v09}             \\
HAT-P-11     & 25.74 &  no &    0.31 & \citet{bakos10}         \\
HD 190360    & 18.12 & yes &    0.24 & \citet{s08}             \\
HD 215497    &  5.40 & yes &    0.23 & \citet{LC10}            \\
\sidehead{\textit{Literature Results for M Stars}}
HD 285968    &  8.42 &  no & $-$0.10 & \citet{e08}             \\
             &       &     &    0.18 & \citet{ja09}            \\
             &       &     &    0.04 & Average                 \\
GJ 436       & 22.88 &  no &    0.02 & \citet{b05}             \\
             &       &     & $-$0.32 & \citet{bean06}          \\
             &       &     &    0.25 & \citet{ja09}            \\
             &       &     & $-$0.02 & Average                 \\
Gl 581       &  1.94 &  no & $-$0.25 & \citet{b05}             \\
             &       &     & $-$0.33 & \citet{bean06}          \\
             &       &     & $-$0.10 & \citet{ja09}            \\
             &       &     & $-$0.23 & Average                 \\
GJ 674       & 11.76 &  no & $-$0.28 & \citet{b07}             \\
             &       &     & $-$0.11 & \citet{ja09}            \\
             &       &     & $-$0.20 & Average                 \\
Gliese 876   &  6.36 & yes & $-$0.03 & \citet{b05}             \\
             &       &     & $-$0.12 & \citet{bean06}          \\
             &       &     &    0.37 & \citet{ja09}            \\
             &       &     &    0.07 & Average                 \\
\enddata
\end{deluxetable}



\begin{thebibliography}{}

 \bibitem[Arenou et al.(1992)]{a92} Arenou, F., Grenon, M., \& G\'omez, A. 1992, \aap, 258, 204

 \bibitem[Asplund et al.(2009)]{a09} Asplund, M., Grevesse, N., Sauval, A. J., \& Scott, P. 2009, \araa, 47, 481

 \bibitem[Bakos et al.(2010)]{bakos10} Bakos, G. \'A., et al. 2010, \apj, 710, 1724

 \bibitem[Bard, Kock \& Kock(1991)]{bkk91} Bard, A., Kock, A., \& Kock, M. 1991, \aap, 248, 315

 \bibitem[Bard \& Kock(1994)]{bk94} Bard, A., \& Kock, M. 1994, \aap, 282, 1014

 \bibitem[Bean et al.(2006)]{bean06} Bean, J. L., Benedict, G. F., Endl, M. 2006, \apj, 653, L65

 \bibitem[Blackwell et al.(1986)]{b86} Blackwell, D. E., Booth, A. J., Haddock, D. J., \& Petford, A. D. 1986, \mnras, 220, 549

 \bibitem[Blackwell et al.(1984)]{b84} Blackwell, D. E., Booth, A. J., \& Petford, A. D. 1984, \aap, 132, 236

 \bibitem[Blackwell et al.(1995)]{b95} Blackwell, D. E., Lynas-Gray, A. E., \& Smith, G. 1995, \aap, 296, 217

 \bibitem[Blackwell et al.(1982a)]{b82a} Blackwell, D. E., Petford, A. D., Shallis, M. J., \& Simmons, G. J. 1982a, \mnras, 199, 43

 \bibitem[Blackwell et al.(1982b)]{b82b} Blackwell, D. E., Petford, A. D., \& Simmons, G. J. 1982b, \mnras, 201, 595

 \bibitem[Bond et al.(2006)]{b06} Bond, J. C., Tinney, C. G., Butler, R. P., Jones, H. R. A., Marcy, G. W., Penny, A. J., \& Carter, B. D. 2006, \mnras, 370, 163

 \bibitem[Bonfils et al.(2007)]{b07} Bonfils, X., et al. 2007, \aap, 474, 293

 \bibitem[Bonfils et al.(2005)]{b05} Bonfils, X., Delfosse, X., Udry, S., Santos, N. C., Forveille, T., S\'egransan, D. 2005, \aap, 442, 635

 \bibitem[Borucki et al.(2010)]{b10} Borucki, W. J., et al. 2010, \apj, 713, L126

 \bibitem[Boss(2010)]{boss10} Boss, A. P. 2010, in IAU Symp. 265, Chemical Abundances in the Universe: Connecting First Stars to Planets, ed. K. Cunha, M. Spite \& B. Barbuy (Cambridge:Cambridge University Press), 391

 \bibitem[Caffau et al.(2010)]{caf10} Caffau, E., Ludwig, H.-G., Steffen, M., Freytag, B., Bonifacio, P. 2010 (arXiv:1003.1190)

 \bibitem[Casagrande et al.(2010)]{c10} Casagrande, L., Ram\'irez, I., Mel\'endez, J., Bessel, M., \& Asplund, M. 2010, arXiv:astro-ph/1001.3142v1

 \bibitem[Castelli \& Kurucz(2004)]{ck04} Castelli, F., \& Kurucz, R. L. 2004, in Proc. IAU Symp. 210, Modelling of Stellar Atmospheres ed. N. Piskunov, et al. (Dordrecht: Kluwer), poster A20 (arXiv:astro-ph/0405087)

 \bibitem[Cutri et al.(2003)]{c03} Cutri, R. M., et al. 2003, VizieR Online Data Catalog, II/246

 \bibitem[da Silva et al.(2006)]{dS06} da Silva, L., et al. 2006, \aap, 458, 609

 \bibitem[Demarque et al.(2004)]{d04} Demarque, P., Woo, J.-H., Kim, Y.-C., \& Yi, S. K. 2004, \apjs, 155, 667

 \bibitem[Endl et al.(2008)]{e08} Endl, M., Cochran, W. D., Wittenmyer, R. A., \& Boss, A. P. 2008, \apj, 673, 1165

 \bibitem[ESA(1997)]{hip97} ESA 1997, The Hipparcos and Tycho Catalogues, SP 1200 (ESA)

 \bibitem[Fischer \& Valenti(2005)]{fv05} Fischer, D. A., \& Valenti, J. 2005, \apj, 622, 1102

 \bibitem[Fuhr, Martin \& Wiese(1988)]{fmw88} Fuhr, J. R., Martin, G. A., \& Wiese, W. L. 1988, J. Phys. Chem. Ref. Data 17, Suppl. 4

 \bibitem[Fulbright et al.(2006)]{f06} Fulbright, J. P., McWilliam, A., \& Rich, R. M. 2006, \apj, 636, 821

 \bibitem[Gehren et al.(2001a)]{g01a} Gehren, T., Butler, K., Mashonkina, L., Reetz, J., \& Shi, J. 2001, \aap, 366, 981

 \bibitem[Gehren et al.(2001b)]{g01b} Gehren, T., Korn, A.J., \& Shi, J. 2001, \aap, 380, 645

 \bibitem[Ghezzi et al.(2009)]{ghezzi09} Ghezzi, L., Cunha, K., Smith, V.V., Margheim, S., Schuler, S., de Ara\'ujo, F.X., \& de la Reza, R. 2009, \apj, 698, 451

 \bibitem[Ghezzi(2010)]{ghezzi10} Ghezzi, L. 2010, Ph.D. thesis, Observat\'orio Nacional, Rio de Janeiro

 \bibitem[Girardi et al.(2002)]{g02} Girardi, L., Bertelli, G., Bressan, A., Chiosi, C., Groenewegen, M. A. T., Marigo, P., Salasnich, B., \& Weiss, A. 2002, \aap, 391, 195

 \bibitem[Gonzalez(1997)]{g97} Gonzalez, G. 1997, \mnras, 285, 403

 \bibitem[Gonzalez(2006)]{g06} Gonzalez, G. 2006, \pasp, 118, 1494

 \bibitem[Gonzalez \& Vanture(1998)]{gv98} Gonzalez, G., \& Vanture, A. D. 1998, \aap, 339, L29

 \bibitem[Hakkila et al.(1997)]{h97} Hakkila, J., Myers, J. M., Stidham, B. J., \& Hartmann, D. H. 1997, \aj, 114, 2043

 \bibitem[H\'ebrard et al.(2010)]{h10} H\'ebrard, G., et al. 2010, \aap, 512, A46

 \bibitem[Holweger et al.(1991)]{h91} Holweger, H., Bard, A., Kock, A., \& Kock, M. 1991, \aap, 249, 545

 \bibitem[Howard et al.(2009)]{h09} Howard, A. W., et al. 2009, \apj, 696, 75

 \bibitem[Ida \& Lin(2004)]{il04} Ida, S., \& Lin, D. N. C. 2004, \apj, 616, 567

 \bibitem[Johnson \& Apps(2009)]{ja09} Johnson, J. A., \& Apps, K. 2009, \apj, 699, 933

 \bibitem[Kaufer et al.(1999)]{kaufer99} Kaufer, A., Stahl, O., Tubbesing, S., N\o{}rregaard, P., Avila, G., Francois, P., Pasquini, L., \& Pizzella, A. 1999, The Messenger, 95, 8

 \bibitem[Kurucz et al.(1984)]{k84} Kurucz, R. L., Furelind, I., Brault, J., \& Testerman, L. 1984, Solar Flux Atlas from 296 to 1300 nm (Cambridge: Harvard Univ. Press)

 \bibitem[Kupka et al.(1999)]{k99} Kupka, F., Piskunov, N., Ryabchikova, T A., Stempels, H. C., \& Weiss, W. W. 1999, \aaps, 138, 119

 \bibitem[Lambert et al.(1996)]{l96} Lambert, D. L., Heath, J. E., Lemke, M., \& Drake, J. 1996, \apjs, 103, 183

 \bibitem[Laws et al.(2003)]{l03} Laws, C., Gonzalez, G., Walker, K. M., Tyagi, S., Dodsworth, J., Snider, K., \& Suntzeff, N. B. 2003, \aj, 125, 2664

 \bibitem[L\'eger et al.(2009)]{l09} L\'eger, A., et al. 2009, \aap, 506, 287

 \bibitem[Lo Curto et al.(2010)]{LC10} Lo Curto, G., et al. 2010, \aap, 512, A48

 \bibitem[Lovis \& Mayor(2007)]{lm07} Lovis, C., \& Mayor, M. 2007, \aap, 472, 657

 \bibitem[Luck \& Heiter(2006)]{lh06} Luck, R. E., \& Heiter, U. 2006, \aj, 131, 3069

 \bibitem[Marcy et al.(2005)]{m05} Marcy, G. W., Butler, R. P., Vogt, S. S., Fischer, D. A., Henry, G. W., Laughlin, G., Wright, J. T., \& Johnson, J. A. 2005, \apj, 619, 570

 \bibitem[May et al.(1974)]{m74} May, M., Richter, J., Wichelmann, J. 1974 , \aaps, 18, 405

 \bibitem[Mayor et al.(2009)]{mayor09} Mayor, M., et al. 2009, \aap, 493, 639

 \bibitem[Mel\'endez et al.(2009)]{m09} Mel\'endez J., Asplund, M., Gustafsson, B., \& Yong, D. 2009, \apj, 704, L66

 \bibitem[Mel\'endez \& Barbuy(2009)]{mb09} Mel\'endez J., \& Barbuy B. 2009, \aap, 497, 611

 \bibitem[Mishenina et al.(2008)]{m08} Mishenina, T. V., Soubiran, C., Bienaym\'e, O., Korotin, S. A., Belik, S. I., Usenko, I. A., \& Kovtyukh, V. V. 2008, \aap, 489, 923

 \bibitem[Mordasini et al.(2009a)]{mordasini09a} Mordasini, C., Alibert, Y., \& Benz, W. 2009, \aap, 501, 1139

 \bibitem[Mordasini et al.(2009b)]{mordasini09b} Mordasini, C., Alibert, Y., Benz, W., \& Naef, D. 2009, \aap, 501, 1161

 \bibitem[O\textquoteright Brian et al.(1991)]{ob91} O\textquoteright Brian, T. R., Wickliffe,  M. E., Lawler, J. E., Whaling, W., Brault J. W., 1991, J. Opt. Soc. Am. B, 8, 1185

 \bibitem[Pasquini et al.(2007)]{p07} Pasquini, L., D\"{o}llinger, M. P., Weiss, A., Girardi, L., Chavero, C., Hatzes, A. P., da Silva, L., Setiawan, J. 2007, \aap, 473, 979

 \bibitem[Perryman et al.(1997)]{p97} Perryman, M. A. C., et al. 1997, \aap, 323, L49

 \bibitem[Pinsonneault et al.(2001)]{p01} Pinsonneault, M. H., DePoy, D. L., \& Coffee, M. 2001, \apj, 556, L59

 \bibitem[Raassen \& Uylings(1998)]{ru98} Raassen,  A. J. J., \& Uylings, P. H. M. 1998, \aap, 340, 300

 \bibitem[Reddy et al.(2003)]{r03} Reddy, B. E., Tomkin, J., Lambert, D. L., \& Allende Prieto, C. 2003, \mnras, 340, 304

 \bibitem[Santos et al.(2001)]{santos01} Santos, N. C., Israelian, G., \& Mayor, M. 2001, \aap, 373, 1019

 \bibitem[Santos et al.(2004)]{s04} Santos, N. C., Israelian, G., \& Mayor, M. 2005, \aap, 415, 1153

 \bibitem[Santos et al.(2005)]{s05} Santos, N. C., Israelian, G., Mayor, M., Bento, J. P., Almeida, P. C., Sousa, S. G., \& Ecuvillon, A. 2005, \aap, 437, 1127

 \bibitem[Smith et al.(2001)]{s01} Smith, V. V., Cunha, K., Lazzaro, D. 2001, \aj, 121, 3207

 \bibitem[Sneden(1973)]{s73} Sneden, C. 1973, PhD thesis, Univ. Texas, Austin

 \bibitem[Sousa et al.(2007)]{s07} Sousa, S. G., Santos, N. C., Israelian, G., Mayor, M., \& Monteiro, M. J. P. F. G. 2008, \aap, 469, 783

 \bibitem[Sousa et al.(2008)]{s08} Sousa, S. G., Santos, N. C., Mayor, M., Udry, S., Casagrande, L., Israelian, G.,  Pepe, F., Queloz, D., \& Monteiro, M. J. P. F. G. 2008, \aap, 487, 373

 \bibitem[Takeda et al.(2005)]{t05} Takeda, Y., Ohkubo, M., Sato, B., Kambe, E., Sadakane, K. 2005, \pasj, 57, 27

 \bibitem[Udry \& Santos(2007)]{us07} Udry, S., \& Santos, N. C., 2007, \araa, 45, 397

 \bibitem[Udry et al.(2006)]{u06} Udry, S., et al. 2006, \aap, 447, 361

 \bibitem[Valenti et al.(2009)]{v09} Valenti, J. A., et al. 2009, \apj, 702, 989

 \bibitem[Valenti \& Fischer(2005)]{vf05} Valenti, J. A., \& Fischer, D. A. 2005, \apjs, 159, 141

 \bibitem[van Leeuwen(2007)]{vL07} van Leeuwen, F. 2007, \apss Library, Vol. 350, Hipparcos, the New Reduction of the Raw Data

 \bibitem[van Leeuwen et al.(1997)]{vL97} van Leeuwen, F., Evans, D. W., Grenon, M., Gro\ss mann, V., Mignard, F., \& Perrymann, M. A. C. 1997, \aap, 323, L61

 \bibitem[Woolf \& Wallerstein(2006)]{ww06} Woolf, V. M., \& Wallerstein, G. 2006, \pasp, 118, 218

 \bibitem[Yi et al.(2003)]{y03} Yi, S. K., Kim, Y.-C., \& Demarque, P. 2003, \apjs, 144, 259

\end{thebibliography}
\end{document}